\documentclass[fleqn,usenatbib]{mnras}
\usepackage[T1]{fontenc}
\usepackage{ae,aecompl}
\usepackage{tabularx}
\usepackage{float}
\usepackage{etoolbox}
\usepackage{color}
\usepackage{natbib}
\usepackage{multirow}
\usepackage{graphicx}	
\usepackage{amsmath}	
\usepackage{amssymb}	
\usepackage[flushleft]{threeparttable}

\newcommand{\Ms}{{\rm M}_\odot}
\newcommand{\bhb}{{\rm BHB}}
\newcommand{\bh}{{\rm BH}}
\newcommand{\gw}{{\rm GW}}
\newcommand{\gc}{{\rm GC}}
\newcommand{\higpus}{\texttt{HiGPUs}}
\newcommand{\phigpu}{\texttt{PhiGPU}}

\title[SMBHs coalescence mediated by massive perturbers]{Supermassive black holes coalescence mediated by massive perturbers: implications for gravitational waves emission and nuclear cluster formation}

\author[M. Arca Sedda et al.]{
Manuel Arca Sedda$^{1}$\thanks{Contact e-mail: \href{mailto:m.arcasedda@ari.uni-heidelberg.de}{m.arcasedda@ari.uni-heidelberg.de}}, 
Peter Berczik$^{1,2,5}$,
Roberto Capuzzo-Dolcetta$^{3}$,
\newauthor
Giacomo Fragione$^{4}$,
Margaryta Sobolenko$^{5}$,
Rainer Spurzem$^{1,2,6}$\thanks{Research Fellow of Frankfurt Institute for Advanced Study}\\
\\
\footnotesize
$^{1}$Zentrum f\"{u}r Astronomie der Universit\"{a}t Heidelberg, Astronomisches Rechen-Institut, M\"{o}nchhofstr. 12-14, D-69120 Heidelberg, Germany\\
$^{2}$National Astronomical Observatories and Key Laboratory of Computational Astrophysics, Chinese Academy of Sciences, 20A Datun Rd., \\Chaoyang District, Beijing 100012, China \\
$^{3}$Dep. of Physics, Sapienza, Universit\'a di Roma, P.le A. Moro 5, 00185 Roma, Italy\\
$^{4}$Hebrew University of Jerusalem Racah Institute of Physics, Givat Ram, 9190401  Jerusalem, Israel\\
$^{5}$Main Astronomical Observatory, National Academy of Sciences of Ukraine, 27 Akademika Zabolotnoho St., 03680 Kyiv, Ukraine \\
$^{6}$Kavli Institute for Astronomy and Astrophysics, Peking University, Yiheyuan Lu 5, Haidian Qu, Beijing 100871, China}

\date{Accepted XXX. Received YYY; in original form ZZZ}

\pubyear{2017}

\begin{document}
\label{firstpage}
\pagerange{\pageref{firstpage}--\pageref{lastpage}}
\maketitle

\begin{abstract}
A large fraction of galactic nuclei is expected to host supermassive black hole binaries (BHB), likely formed during the early phase of galaxies assembly and merging. In this paper, we use a large set of state-of-art numerical models to investigate the interplay between a BHB and a massive star cluster (GCs) driven toward the galactic centre by dynamical friction. Varying the BHB mass and mass ratio and the GC orbit, we show that the reciprocal feedback exerted between GCs and the BHB shapes their global properties. We show that, at GC-to-BHB mass ratios above 0.1, the GC affects notably the BHB orbital evolution, possibly boosting its coalescence. This effect is maximized if the GC moves on a retrograde orbit, and for a non-equal mass BHB. We show that the GC debris dispersed around the galactic nucleus can lead to the formation of a nuclear cluster, depending on the BHB tidal field, and that the distribution of compact remnants resulting from the GC disruption can carry information about the BHB orbital properties. We find that red giant stars delivered by the spiralling GC can be disrupted at a rate of $\simeq (0.7-7)\times 10^{-7}$ yr$^{-1}$ for BHB masses $\sim 10^7\Ms$, while tens to hundreds of stars can be possibly observed in the galactic halo as high-velocity stars, with velocities up to $\sim 2000$ km s$^{-1}$, depending on the BHB orbital properties.
\end{abstract}


\section{Introduction}
Most of the observed galactic nuclei over the whole Hubble sequence harbour super-massive black holes (SMBHs) with masses in the range $10^{6}-10^{10}\Ms$  \citep{ferr00}, likely forming at high redshift a few (1-2) Gyr after the Big Bang \citep{Kormendy13}.
Uncovering the processes that regulate the formation of SMBHs can allow us to shed light on galaxy formation and evolution.
According to the standard $\Lambda$ cold-dark matter ($\Lambda$CDM) cosmological theory, galaxies assembly in a hierarchical fashion, undergoing several mergers with each other over their whole life-time \citep{volonteri03}. When two colliding galaxies host at their centre an SMBH, their subsequent evolution inevitably leads to the formation of a super-massive black hole binary \citep[BHB,][]{begelman80}.
Another common feature of galaxies across the entire Hubble sequence is the presence, in their centre, of a massive and compact stellar cluster referred to as nuclear cluster (NSC). 
NSCs are the densest stellar systems observed so far, with typical masses of $\approx 10^6-10^8\Ms$ and half-mass radii of $\approx 2-5$ pc \citep{boker02,cote06,Turetal12,georgiev14}. NSCs are made up of a predominant old stellar population ($\tau>1$ Gyr) and show also the presence of a young stellar population ($\tau<100$ Myr) \citep{rossa}.

As unveiled by observations over the last decade, the main properties of NSCs and their hosts tightly correlate, although NSCs scaling relations differ significantly from those of SMBHs \citep{scot,LGH,ERWGD,georgiev16,melo16}, thus suggesting that NSCs and SMBHs likely form following different path. 
In some cases, at galaxies masses in between (masses $10^{10}-10^{11}\Ms$), NSCs and SMBHs co-exist in the same nucleus \citep{graham09}, making worth the study of their interplay.

According to the so-called {\it dry-merger} scenario, NSCs form via repeated mergers of massive star clusters (GCs) that segregate toward the inner galactic regions due to dynamical friction \citep{Trem75,Dolc93}. This mechanism seems to explain excellently NSCs structural properties \citep{DoMioA,DoMioB,AMB,perets14,tsatsi17,abbate17,ASK17,frak18,frlgk18,fragione17b} and their observational scaling relations \citep{antonini13,gnedin14,ASCD14b}. The dry-merger scenario is very general, providing a suitable explanation for the formation of NSCs in dwarf galaxies \citep{ASCD16b,ASCD16b} and in galaxies harbouring an SMBH \citep{ASCD15He}, and for the observational absence of NSCs in massive ellipticals \citep{ASCDS16,ASCD17}.


In the context of galaxies hosting an BHB, the dry-merger scenario is still poorly investigated. This is due to the extreme computational complexity hidden behind the physical problem, since it involves different space and time scales, from the BHB pc scale to the kpc scale of the host galaxy. Recently, the overwhelming increase in the computational capacity of graphic processing units (GPUs) and the development of dedicated software to maximize their use, lead to the possibility to integrate such a problem keeping a good compromise between the computational cost and the reliability of the model.

BHBs are believed to undergo three main stages during their lifetime. Soon after their parent galaxies merge, they sink independently due to dynamical friction \citep{Cha43I} in the centre of the newborn galaxy, eventually leading to the formation of a binary system on a dynamical friction time-scale. Once in the galactic nucleus, the binary starts loosing energy and angular momentum due to three-body slingshots of stars passing nearby. The duration of this stage is still highly uncertain and depends on the loss-cone refill of the scattered stars, and the BHB binary may stall at parsec scale \citep{mikkola92,quin96,sesana06}. The last stage is characterized by the energy loss due to gravitational radiation and the BHB will merge within the \citet{peters64} timescale.
Several mechanisms have been proposed to solve the so-called "final parsec problem", as a triaxial morphology of the environment surrounding the BHB \citep{berczik06,khan12,vasiliev15,khan16,bortolas16,dosopoulou17}.

There are a number of reasons for which the dry-merger scenario for BHBs is a worth problem to be investigated. For instance, during the MBHB evolution, GCs can undergo several close fly-by which can have crucial implications on the evolution of both the GC and the MBHB. As first discussed in \citet{fragione16}, a GC scattering over a MBHB can lead to the production of high-velocity stars with velocities exceeding $\sim 1000$ km s$^{-1}$, a phenomenon similar to strong GC-SMBH interactions investigated in earlier studies \citep{ASCDS16,cd15,ASCD17c,fragione17}.

Hence, repeated star cluster infalls can significantly affect the morphology of the BHB surroundings possibly altering  its own dynamics and enhancing its hardening rate. 
Two different class of interactions can turn the binary onto a new hardening phase: an impulsive interaction, which develops during the cluster passage at pericentre and extract energy and momentum from the BHB, and a secular interactions, which is exerted from the cluster debris deposited around the BHB over time. Both the mechanisms can push the BHB towards a hardening phase, which may accelerate the BHB coalescence together with other factors, such as the presence of a gaseous disc around the binary \citep{cuadra09,roe11,roe14,armit15}.

A similar process can work also if the BHB is surrounded by gaseous clumps or molecular clouds. \cite{goicovic17} modelled the evolution of a gas cloud impacting a circular BHB at varying impact parameters, inferring the BHB behaviour in the case of gas accretion by one or both the SMBHs, the effect of multiple infalling clouds and the role of the non-accreted material on the BHB long-term evolution. They found that even in the regime of small cloud-to-BHB mass ratio, $0.01$, repeated gas clouds can drive the BHB toward the GW-dominated regime, leading the BHB to merge in reasonable time-scales. Expanding their previous setup, \cite{goicovic18} investigated the effect of multiple clouds impacting a circular BHB, showing that matter accretion and angular momentum transfer dominates the BHB evolution.

In this paper, we make use of direct $N$-body simulations to investigate how repeated scatterings of a massive star cluster over an BHB may affect the evolution of the binary and the galactic nucleus. 
We discuss the impact of GC flybys on the BHB, the possible formation of a NSC around the BHB and the distribution of compact remnants delivered from the infalling GC to the galactic centre. Varying BHB total mass and mass ratio and its orbital properties, the GC mass and orbits and the mutual configuration between the BHB and GC orbital planes, we can constrain the parameters space that can results in the BHB merging or the formation of a NSC.
Our paper enlarges and generalizes results of a recent paper by \cite{bortolas17} on a similar topic.
The long-term evolution of BHBs is a subject of crucial importance for improving our knowledge of galaxy formation processes and, also, for consolidating the basis needed to study BHB mergers with the next generation of GW detectors. As a long-standing problem, BHB evolution was mostly studied from two different perspectives: stellar dynamics and gaseous dynamics. 
With regards to the topics explored in this work, for instance, several papers have been developed recently on the effect of gaseous clumps and clouds on the evolution of the BHB \citep{goicovic16,goicovic17,maureira18,goicovic18}. 
Similarly, this paper focuses on the effect of stellar systems on the BHB evolution, thus representing a perfect counterpart to the work based on gaseous dynamics. Comparing these works would allow to improve our knowledge of the various roles played by gas and stars in the evolution of massive BHBs.

The paper is organized as follows. In Section 2, we present the numerical method we use to handle the problem. In Section 3, we present our results. In Section 4, we discuss the implication of our results. Finally, we draw our conclusions in Section 5.

\section{Numerical method}

\subsection{Initial conditions setup}
In this paper we investigate the evolution of a BHB subjected to the perturbation of a massive GC. We took into account the tidal field of the background galaxy, treating it as an external static potential. The star cluster is modelled according to a \cite{King} profile, while the total number of particles is limited to $2^{16}\simeq 65$k. The parameter space associated to such a problem is vast. In fact, the BHB is characterised by at least four parameters: binary mass $M_\bhb$ and mass ratio $q_\bhb$, semi-major axis $a_\bhb$ and eccentricity $e_\bhb$. The GC, instead, can be constrained through its potential well $W_0$ and total mass $M$. According to the \cite{King} models, fixing $W_0$ corresponds to fixing the cluster concentration parameter $c=r_c/r_t$, given by the ratio between the core ($r_c$) and tidal ($r_t$) radius. On the other hand, the tidal radius depends on the cluster orbit through the relation

\begin{equation}
r_t = \left(\frac{GM}{\Omega^2-{\rm d}^2\Phi/{\rm d}r^2}\right)^{1/3},
\label{rtidal}
\end{equation}

where $\Omega$ is the GC angular velocity, and ${\rm d}^2\Phi/{\rm d}r^2$ the second order derivative of the background gravitational field\footnote{Note that Equation \ref{rtidal} is valid under the assumption of spherical symmetry for the background gravitational potential}.
In the simplest approximation that the BHB separation is much smaller than its distance from the GC and that it dominates the dynamics, the equation reduces to 
\begin{equation}
r_t = r_p \left(\frac{M}{3M_\bhb}\right)^{1/3},
\label{rtidalt}
\end{equation}
being $r_p$ the GC periastron. Although Equation \ref{rtidalt} was originally derived for point-like mass moving on circular orbits, it approximate reasonably well the tidal radius of an extended object moving on an eccentric orbit during the passage at periastron \citep{king62,ASCDS16}.
Once the GC orbit is assigned, the tidal radius can be calculated and combined with the value of $W_0$ chosen in order to get $r_c$.

Another degree of freedom is represented by the GC pericentre ($r_p$) and eccentricity ($e$), as well as the orbit inclination with respect to the BHB orbital plane, $i$, and the mutual orientation of the two orbital ascending nodes $\omega$. Note that the definition of the inclination angle $i$ is as the one between the BHB and the GC orbital angular momentum vectors, where $i=0$ corresponds to prograde and $i=\pi$ to retrograde configurations.

Finally, the galaxy background can significantly alter the first part of the GC orbital evolution. According to the widely used \cite{Deh93} models, the galactic density profile can be described as 
\begin{equation}
\rho(r)=\frac{(3-\gamma)M_g}{4\pi r_g^3}\left(\frac{r}{r_g}\right)^{-\gamma}\left(1+\frac{r}{r_g}\right)^{-4+\gamma},
\label{dens}
\end{equation}
where $M_g$, $r_g$ represent the galaxy mass and length scale, respectively, whereas $\gamma$ represents the inner slope of the density profile.

Therefore, a lower limit to the total number of parameters needed to characterize such a system is given by the combination of the BHB parameters (4), the GC structural and orbital properties (6) and the galaxy parameters (3) , namely:
\begin{itemize}
\item BHB parameters: semi-major axis $a_\bhb$, eccentricity $e_\bhb$, total mass $M_\bhb$ and mass ratio $q_\bhb$;
\item GC structural parameters: Mass $M$, potential well $W_0$, concentration $c$;
\item GC orbital parameters: semi-major axis $a$, eccentricity $e$ and pericentre $r_p$;
\item galaxy parameters: galaxy mass $M_g$, typical radius $r_g$ and density profile slope $\gamma$.
\end{itemize}

Assuming only 3 values for each parameter would imply then a total number of $\sim 1.6\times 10^6$ different models.

In order to keep the number of models reasonable in terms of computational request and human times, we limited the number of degrees of freedom in our problem as follows:
\begin{itemize}
\item we assumed an equal mass BHB and kept fixed its apocentre;
\item we fixed the GC $W_0$, $M$ and $r_p$;
\item we assumed the GC and BHB orbits in a coplanar, counter-rotating configuration;
\item we fixed the galaxy background.
\end{itemize}

This choices reduced our problem to only three parameters: the BHB mass $M_\bhb$ and its eccentricity, $e_\bhb$, and the GC eccentricity $e$.

We assumed two different values for $M_\bhb = 10^7 - 10^8 \Ms$, three values for $e_\bhb = 0, 0.5, 0.9$ and two for $e = 0.5, 0.9$, thus implying 12 different sets of initial conditions.
In addition, we ran three models in which the GC and the BHB orbital planes are in a co-rotating configuration ($i=0$), and one in which the binary mass ratio is $q_\bhb = 0.1$, namely $M_{{\rm BH},1} = 10^7\Ms$ and $M_{{\rm BH},2} = 10^6\Ms$.

The GC density profile is described by a King model with $W_0=6$ and core radius $r_c=0.24$. The BHB initial apocentre is fixed at 1 pc, while the GC orbit is kept in such a way that its pericentre is twice the binary separation, $2$ pc.

The galaxy model is described by a \citet{Deh93} model with total mass $M_g=10^{11}\Ms$, length scale $r_g=2$ kpc and inner slope of the density profile $\gamma = 0.5$.

To further explore the limits imposed by our choices, we ran a set of additional models, which are described in detail in the next section.

\subsection{Limits and advantages of the simulated sample}

The assumptions made in the present study aim at representing a good compromise between the large number of simulations that would be needed to cover the phase-space, and the minimum number of models required to obtain a handful set of results that possibly constitute the basis for future investigations. 

The assumption of keeping the GC pericentre fixed to $r_p = 2a$ puts us in between two different regimes: a ``catastrophic'' regime $r_p \lesssim a$, in which the spiralling cluster can hit one of the BHB components, and a ``secular'' regime $r_p\gg a$, in which the GC perturbations act secularly on the BHB dynamics, while the GC spiral toward the centre due to dynamical friction.
Both these regimes offer interesting perspectives to study. In the catastrophic regime, the GC net contribute to the BHB evolution will be a combination of the tidal force exerted from the GC on the BHB when it moves far at its apocentre, and the outcomes of possible GC-BH collisions. In the secular regime, instead, the GC can induce secular perturbations on the BHB evolution, approaching slowly the inner galactic regions. The time varying tidal field acting on the BHB can drive an increase in its eccentricity, similarly to what is expected from a third SMBH falling toward a BHB after a galaxy merger event \citep[see for instance][]{bonetti18}.

A proper modelling of the secular regime should include a treatment for the dynamical friction acting on the GC. As discussed deeply in earlier papers, this would require the simultaneous use of at least $1$ million particles, in order to obtain a reliable number of particles in both the cluster and the galaxy \citep{ASCD15He,ASCD17}. Since direct N-body computational time scales roughly as $N^2$, using 1 M particles to perform simulations in the secular regime and taking advantage of the available hardware would increase the time needed for the simulation to be accomplished by a factor $\sim 256$. 
Due to this, we leave the discussion of this regime to a future work. 

Investigating the catastrophic regime seems more affordable with our current setup, although a detailed comparison with all the models would require to double the number of simulations to be performed. In order to balance the demanding of computational resources and the need for a solid simulation database, we run a further model assuming $r_p=1$, keeping as a reference case one of the most interesting, as detailed below.

In all the simulations performed we kept fixed the GC main parameters: mass, potential well and core radius. Regarding the cluster properties, we note that $W_0$, $r_c$ and $r_p$ are inherently connected due to the choice of a \cite{King} model to represent the cluster.

The population of Galactic GCs is characterized by a distribution of ${\rm Log} ~c$ peaking at $1.25-1.50$ \citep[][updated in 2010]{harris96}, which corresponds to typical $W_0\simeq 6-7$. Hence, the choice $W_0=6$ would represent a reliable GC model based on our current understanding of Milky Way GCs. Since $W_0$ is uniquely connected to $c$, setting $W_0 = 6$ implies $c = r_t/r_c = 18$. 

According to Equation \ref{rtidal}, the tidal radius at the GC pericentre is $r_t = 0.30-0.65$ pc, with larger values corresponding to a smaller BHB mass. Hence, to have a GC model filling its Roche lobe during the pericentre passage we should set $r_c = r_t/18 = 10^{-2}$ pc, a value that hardly matches with the observed core radius distribution for galactic GCs. In order to find a good compromise between observational limits and reliability of the model, we assume $r_c = 0.24$ pc. This ensures that $r_c$ is contained within the GC Roche lobe during the passage at pericentre.

A smaller core radius would be unreliable from an observational point of view, yet a viable alternative to our model could be assuming a smaller concentration. The minimum $c$ value corresponds to $W_0=2.5$, and implies $r_t = 3.891 r_c$ \citep{King}, which means a GC core radius $r_c \simeq 0.07-0.17$ pc, according to our galaxy model. 
Therefore, to provide a study based on underfilling or filling GC models seem hardly compatible with observational constraints. 

In a forthcoming paper, we will try to investigate the role of GC structural parameters in determining the BHB evolution, modelling a sample of GC characterized by different properties and moving simultaneously around the galactic centre. 

Regarding the assumed GC mass, we note that clusters lighter than $5\times 10^5\Ms$ are expected to contribute poorly to the galactic nucleus evolution, as they are most likely subject to tidal disruption, as shown in several numerical and semi-analytical studies \citep{ASCD15He,ASCD16b,ASCD17}. The effect of heavier clusters is partly addressed in our simulations, which cover the BHB-to-GC mass ratio 10-100, thus providing a general idea of the role played by the GC mass.

Our current set of simulations focuses mostly on retrograde configurations (inclination angle $i = 180^\circ$) but in a few cases, where we provided a direct comparison with the corresponding prograde systems ($i = 0^\circ$). Results coming from intermediate cases are not automatically limited between these two extremes, thus representing a region of the phase space that deserves to be investigated in more detail. In order to provide a short view on the outcomes of off-plane configuration, we perform two additional runs assuming $r_p = 1-2$ pc and $i=90^\circ$. 

Finally, we run a further model characterized by a BHB total mass $M=1.05\times 10^8\Ms$ and mass ratio $q=0.05$. This system has properties similar to those of the BHB that can, in principle, form in consequence of a merger between the Milky Way and Andromeda. 
Although representing a toy model of the Milky Way - Andromeda future, it provides an outlook on the effects of a GC impacting a binary in which one of the components has a comparable mass, while the other is much more massive.
We called this model Milkomeda, following earlier studies on the subject \citep{Cox07}.

The models main parameters are summarized in Table \ref{tab1}. We divided the among ``basic'' and ``supporting''. Basic models aim at providing a general view on the problem, while supporting models serve as comparison beds, to better highlight the role of the initial conditions in the BHB-GC evolution.

Since a full exploration of the whole parameter space would require a very large amount of 
computational resources, our study would serve as the basis to develop a handle of 
simulations, modelling many GCs orbiting the BHB, characterized by initial conditions draw to explore the regions of the phase space where the effect on the BHB is maximized and minimized. 

This will be the base of a forthcoming investigation conducted in the footsteps of our current results. Also, the results discussed here could serve as a reference to be compared with works focusing on the interactions between a BHB and several gaseous clouds, like recently studied by \cite{goicovic17}.

\begin{table*}
\begin{center}
\caption{Models parameters}
\begin{tabular}{lccccccccc}
\hline
ID & $M_\bhb$ & $q_\bhb$ & $e_\bhb$ & $a_\bhb$ & $M_\gc/M_\bhb$ & $r_\gc$ & $e_\gc$ & $i$ & $T_\bhb$\\
   & $10^7\Ms$ & & & pc &  & pc & & $^{\circ}$ & $10^3$ yr \\
\hline
\multicolumn{10}{c}{{\bf Basic models}}\\
\hline
1  & $1$   & $1$      & $0$      & $1$     & $10^{-1}$  & $5.5$  & $0.5$ & $180$ & $4.7$\\
2  & $1$   & $1$      & $0$      & $1$     & $10^{-1}$  & $37.5$ & $0.9$ & $180$ & $4.7$\\
3  & $1$   & $1$      & $0.5$    & $0.67$  & $10^{-1}$  & $5.5$  & $0.5$ & $180$ & $2.6$\\
4  & $1$   & $1$      & $0.5$    & $0.67$  & $10^{-1}$  & $37.5$ & $0.9$ & $180$ & $2.6$\\
5  & $1$   & $1$      & $0.9$    & $0.53$  & $10^{-1}$  & $5.5$  & $0.5$ & $180$ & $1.8$\\
6  & $1$   & $1$      & $0.9$    & $0.53$  & $10^{-1}$  & $5.5$  & $0.5$ & $0$   & $1.8$\\
7  & $1$   & $1$      & $0.9$    & $0.53$  & $10^{-1}$  & $37.5$ & $0.9$ & $180$ & $1.8$\\
8  & $1.1$   & $0.1$     & $0$     & $1$     & $10^{-1}$  & $5.5$ & $0.5$ & $180$  & $4.5$\\
9  & $10$   & $1$      & $0$       & $1$     & $10^{-2}$  & $5.5$  & $0.5$ & $180$ & $1.5$\\
10 & $10$   & $1$      & $0$       & $1$     & $10^{-2}$  & $37.5$ & $0.9$ & $180$ & $1.5$\\
11 & $10$   & $1$      & $0$       & $1$     & $10^{-2}$  & $37.5$ & $0.9$ & $0$   & $1.5$\\
12 & $10$   & $1$      & $0.5$     & $0.67$  & $10^{-2}$  & $5.5$  & $0.5$ & $180$ & $0.8$\\
13 & $10$   & $1$      & $0.5$     & $0.67$  & $10^{-2}$  & $5.5$  & $0.5$ & $0$   & $0.8$\\
14 & $10$   & $1$      & $0.5$     & $0.67$  & $10^{-2}$  & $37.5$ & $0.9$ & $180$ & $0.8$\\
15 & $10$   & $1$      & $0.9$     & $0.53$  & $10^{-2}$  & $5.5$  & $0.5$ & $180$ & $0.6$\\
16 & $10$   & $1$      & $0.9$     & $0.53$  & $10^{-2}$  & $37.5$ & $0.9$ & $180$ & $0.6$\\
\hline
\multicolumn{10}{c}{{\bf Supporting models}}\\
\hline
17$^{1}$           & $10.5$ & $0.05$  &$0.5$ & $0.67$  & $10^{-2}$  & $5.5$  & $0.5$  & $180$ & $0.8 $ \\
18                 & $1 $   & $1$   &$0.5$ & $0.67$  & $10^{-1}$  & $37.5$ & $0.9$  & $ 90$ & $2.6 $ \\
19                 & $1 $   & $1$   &$0.5$ & $0.67$  & $10^{-1}$  & $19$   & $0.9$  & $180$ & $2.6 $ \\
20                & $1 $   & $1$   &$0.5$ & $0.67$  & $10^{-1}$  & $19$   & $0.9$  & $ 90$ & $2.6 $ \\
\multirow{2}{*}{21$^{2}$}  & \multirow{2}{*}{$1 $}&\multirow{2}{*}{$1.0$}&\multirow{2}{*}{$0.5$}& \multirow{2}{*}{$0.67$}  & $10^{-1}$  & $5.5$  & $0.5$  & $180$ & \multirow{2}{*}{$2.6$} \\
                           &     &     &     &         & $10^{-1}$  & $37.5$ & $0.9$  & $180$ & \\
\hline
\end{tabular}
\label{tab1}
\end{center}
\begin{tablenotes}
\item Col. 1: model identification number. Col. 2-5: BHB binary mass, mass ratio, eccentricity and semi-major axis. Col. 6-8: GC mass, apocentre and eccentricity. Col. 9: GC orbital inclination, setting the BHB orbital plane as a reference frame. Col. 10: BHB initial period. 
\item $^{1}$ Throughout the text we refer to this model as Milkomeda, since it is composed of two BHs similar to those hosted in the Milky Way and Andromeda galaxies.
\item $^{2}$ Simulation 21 contains two GCs, whose properties are indicated in the table in two different rows.
\end{tablenotes}
\end{table*}

\subsection{Numerical strategy and convergence tests}

The simulations have been carried out using both the \higpus~ \citep{Spera} and \phigpu ~\citep{berczik11} direct $N$-body codes, based on Hermite's integration schemes ($6^{th}$ and $4^{th}$ order, respectively) and tailored to fully exploit the advantages of graphic processing unit computing.
We performed several test-runs to compare the codes, finding a remarkably good agreement between the outputs and similar computational resources consumption. 

\begin{figure}
\centering
\includegraphics[width=8cm]{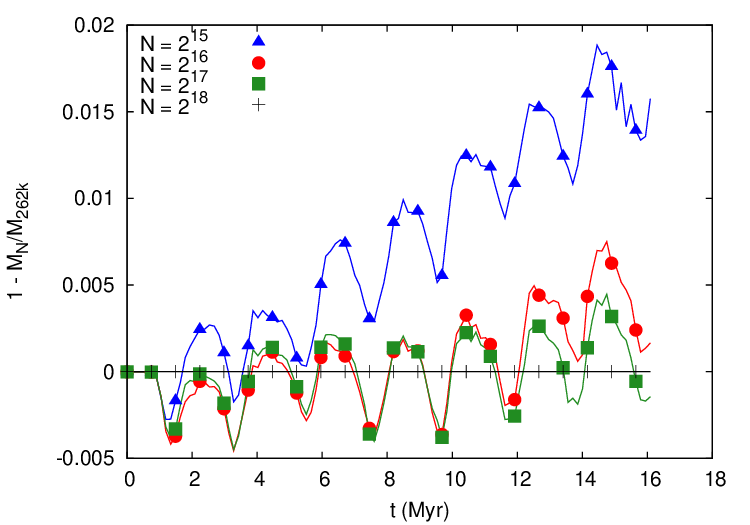}
\caption{
Relative error committed in evaluating the GC mass at different $N$, $1-M_N/M_{262{\rm k}}$. We used model with $N=262$k as a reference.}
\label{mlos}
\end{figure}

In order to investigate possible spurious effects on the GC orbit related to the relatively low number of particles used to model it, we ran some tests at varying particle numbers up to $262$ k particles.
Figure \ref{mlos} shows the relative error committed in evaluating 
the GC mass when decreasing the total number of particle, assuming as a comparison testbed the model with $N=262$k, i.e. $1-M_N/M_{262{\rm k}}$. It appears evident that above $2^{16} = 65536$ particles the results tend to converge, being characterized by a relative error below $1\%$. Hence, we decide to run our models using $N = 65536$, which guarantees the best balance between computational load and simulation accuracy.

The recent advancement in direct $N$-body modelling recently allowed to achieve the ``1 million particle'' goal in the field of GC modelling \citep{wang16} and galactic nuclei \citep{ASCD15He,ASCD16b,ASCD17}. However, the problem investigated here differs from earlier works due to the multiple physical scales involved in the evolution, that must be treated at the same time. This translates in an enhanced computational complexity dictated by the evolution of the subsystem characterized by the smaller time-scale, i.e. the BHB.
We carried out our simulations up to 15 Myr, a time much larger than the maximum BHB orbital period in our models. 

We took advantage of ASTROC16a and ASTROC16b, two high-performance computing workstation at the University of Rome Sapienza Physics department, hosting 4 GPU Nvidia Titan X each, the Kepler cluster, composed of 12 nodes hosting an Nvidia K20 card and the Jureca cluster, a large computational facility comprising up to 74 nodes hosting 2 Nvidia K40 cards each, provided by the J\"ulich computing centre.

Each simulation requires roughly 80 hours to be carried out on a single Nvidia 40K; this immediately outlines the importance of using a relatively low N to model the cluster, and the importance of using GPUs.

\begin{figure*}
\centering
\includegraphics[width=8cm]{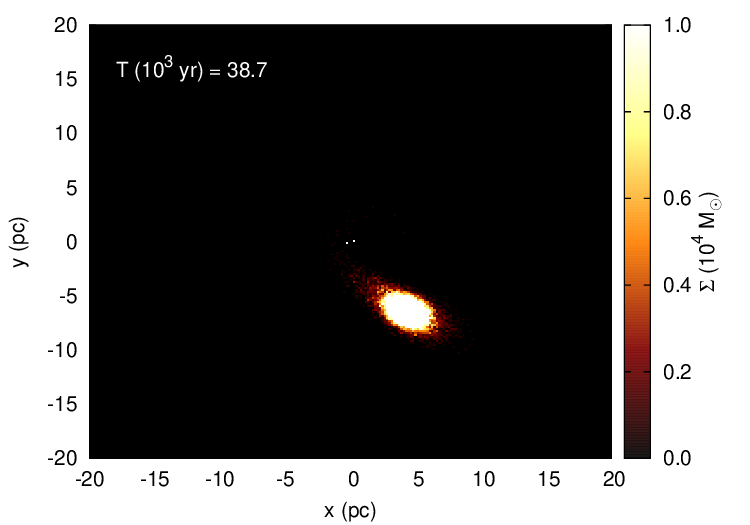}
\includegraphics[width=8cm]{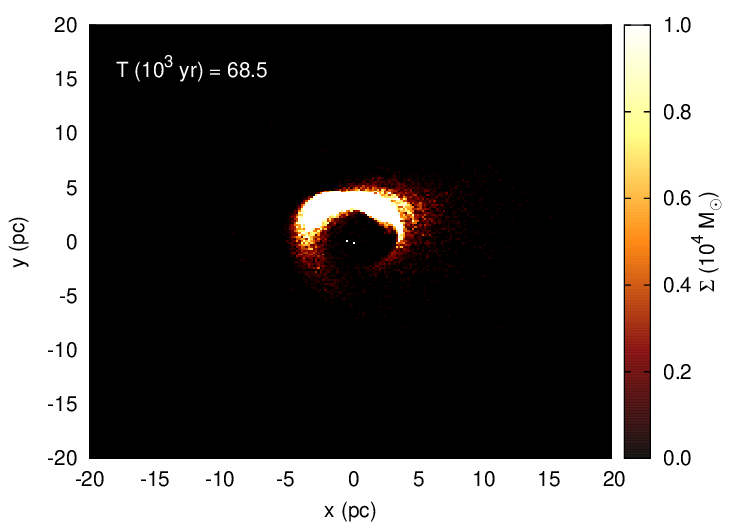}\\
\includegraphics[width=8cm]{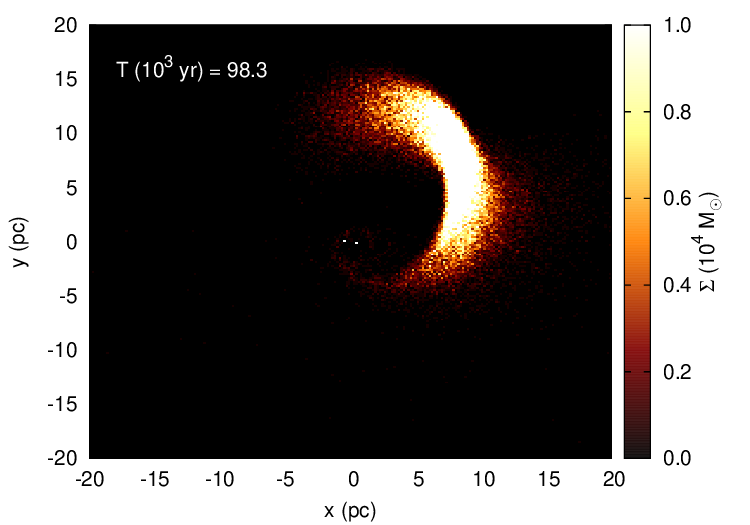}
\includegraphics[width=8cm]{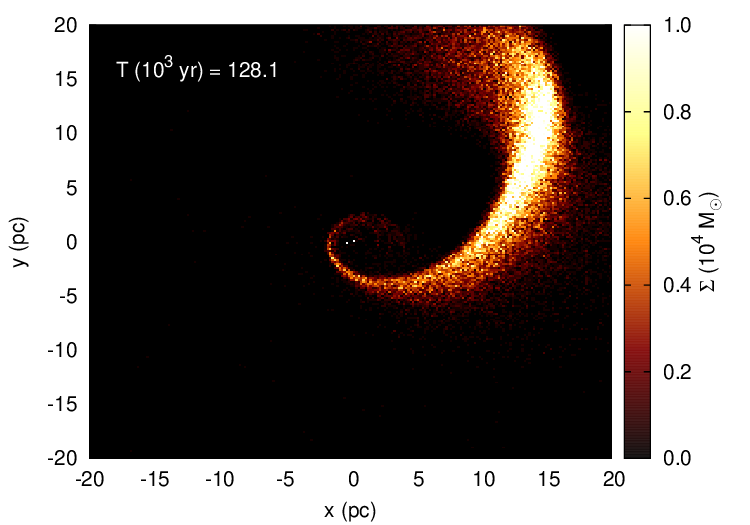}
\caption{Snapshots of one of our models (ID 14). The two white spots represent the SMBHs. The surface density color-coded map refers to $\Ms$ pc$^{-2}$.}
\label{snap}
\end{figure*}

\section{Results}

In this Section, we present the main outcomes of our simulations, discussing the effects of GC flybys on BHB evolution in Section \ref{BHBMER}, the production of high- and hypervelocity stars (Section \ref{hyperS}), and the possible implications for NSC formation in Section \ref{ncform}. 

In general, our simulations suggest that the passing-by GC significantly alters the BHB evolution if its mass is at least 0.1 times the BHB mass. The GC-BHB orbital orientation plays an important role in determining the BHB late evolution. We find that the BHB hardening is more efficient in prograde configurations, while in retrograde the GC-BHB interactions drive a remarkable increase of the BHB eccentricity. The resulting effect is a significant reduction of the BHB merging time-scale. During the GC-BHB close encounter, a number of stars are ejected away from the galactic nucleus and the galaxy, reaching velocities up to $10^4$ km s$^{-1}$. These stars, which may appear observationally as high- and hypervelocity stars, can be used to obtain information about the galaxy nucleus properties. For instance, we find that HVSs coming from retrograde models are characterized by a broader velocity distribution, with a high-end tail longer than prograde ones.
The GC debris accumulates around the BHB, leading to the formation of an over-density that can be observed as an NSC, or a nuclear disc-like structure. The GC mass deposited around the BHB correlate weakly with the GC and the BHB orbital parameters, being larger at high inclinations, i.e. retrograde models, and at larger ratios between the BHB pericentre and the GC apocentre.

\subsection{The impact of GCs flyby on the BHB evolution}
\label{BHBMER}
In this section we focus the attention on the perturbations that the infalling GC induces on the BHB.
Figure \ref{snap} shows several snapshots of one of the performed simulations, making evident the GC disruption shortly after its passage at pericentre. Several stars bind to the BHB and
populate the loss-cone, thus extracting energy and angular momentum from the binary and turning on a hardening phase.

An isolated binary composed of point-like objects in circular orbit, which loses energy and angular momentum through GWs emission undergoes coalescence on a time-scale \citep{peters64}
\begin{equation}
t_\gw \simeq \frac{5}{256}\frac{c^5 a_\bhb^4}{G^3m_1m_2(m_1+m_2)}.
\label{peters}
\end{equation}
Equation \ref{peters} can also be used to estimate the merger time for highly eccentric binaries, provided that $t_\gw$ is multiplied by $\sim 1.8(1-e^2)^{7/2}$.
Therefore, a significant reduction of the BHB semi-major axis, or the increase of its eccentricity can reduce significantly the amount of time needed for the SMBHs to merge.

\begin{figure*}
\includegraphics[width=8cm]{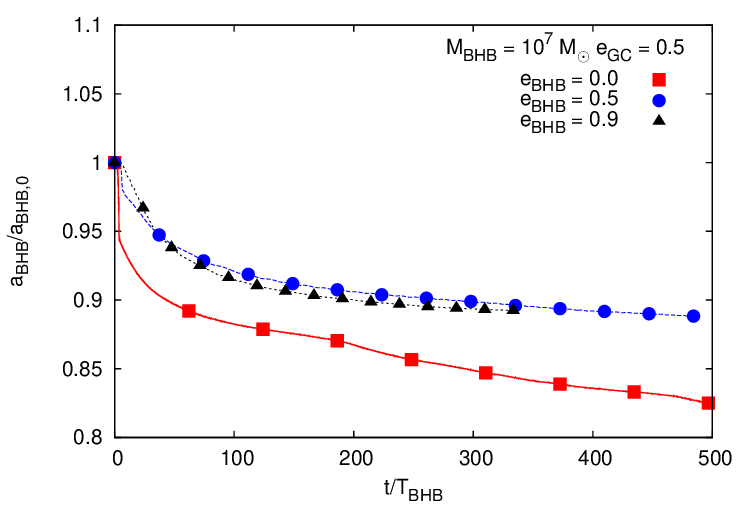}
\includegraphics[width=8cm]{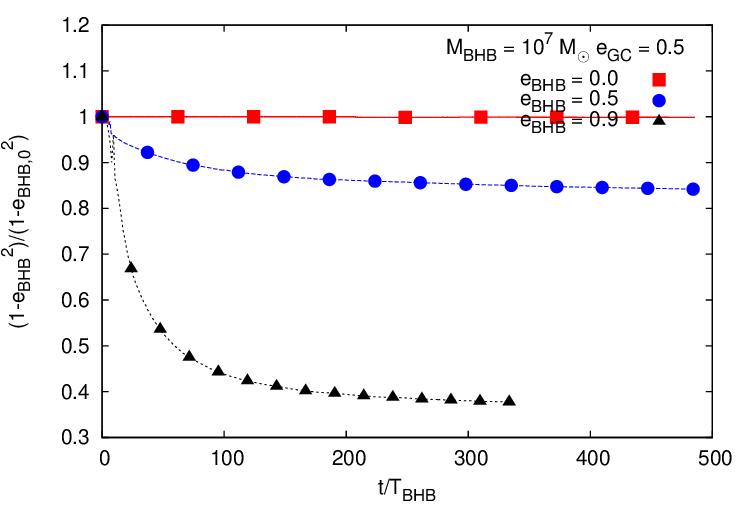}\\
\includegraphics[width=8cm]{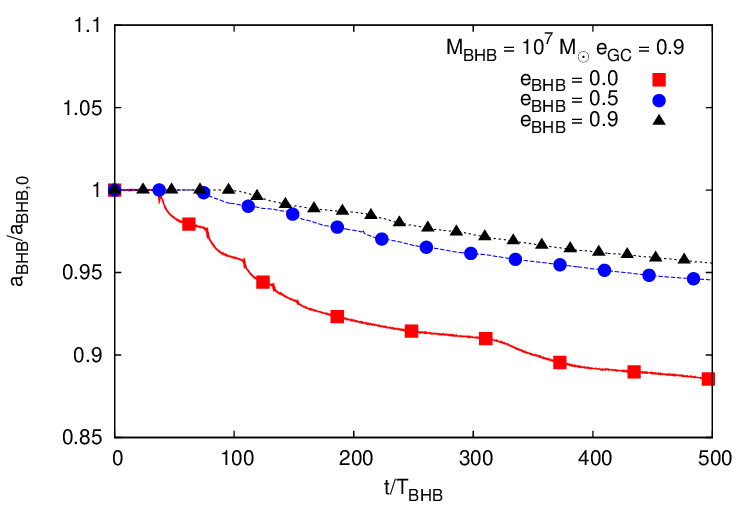}
\includegraphics[width=8cm]{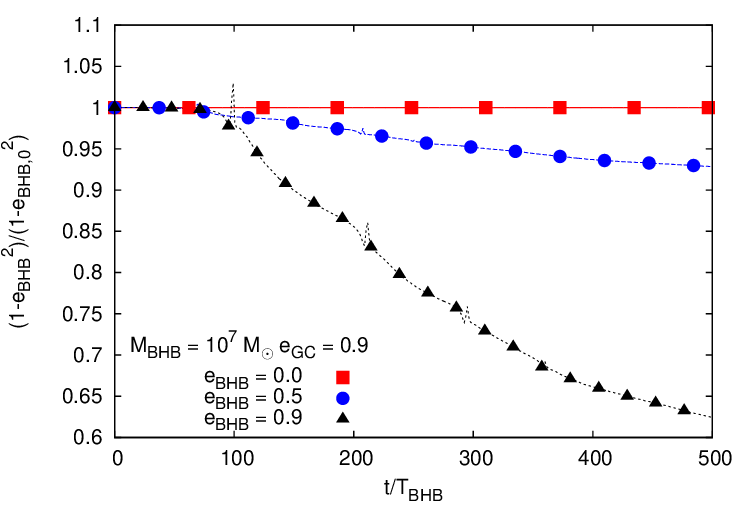}\\
\caption{Time evolution of the BHB semi-major axis in 6 of the models investigated (left panels) and the corresponding time evolution of the ratio $1-e_\bhb^2$, normalized to its initial value (right panels). Top panels refer to models with $e_\gc = 0.5$, while bottom panels are characterized by $e_\gc = 0.9$. The time is normalized to the BHB initial orbital period, $T_\bhb$.}
\label{f2}
\end{figure*}

In all the investigated models, the GC impinges a non-negligible acceleration on the binary, causing its hardening and, in some cases, leading to an excitation of the binary orbital eccentricity. 
Figure \ref{f2} shows the time evolution of $a_\bhb$ and the quantity $1-e_\bhb^2$ for some of the models characterized by $M_\gc/M_\bhb = 0.1$.
We note that the dimensionless squared specific angular momentum $1-e_\bhb^2$ can be used to quantify the variation of the binary coalescence time-scale.

The figure outlines at a glance two important information: i) the GC actually affects the binary evolution, and ii) the efficiency of the GC-SMBH interaction depends on both the GC and BHB eccentricities.

In the case of a zero-eccentricity BHB, we found that the factor $1-e_\bhb^2$ is not significantly affected by the GC motion, quite independently on the GC orbital eccentricity. 
Our findings find a good agreement with earlier works based on three-body scattering experiments. In the limit $M_\bhb\gg m_*$, scattered stars can be considered as massless perturbers, and multiple BHB-stars interactions can be treated as a series of restricted three-body interactions. Under this approximation, \cite{quinlan96} showed that a BHB having $e_\bhb = 0$ tends to remain circular \citep[see also][]{sesana06}, a result that our direct N-body simulations seem to support.

Differently, when $e_\bhb > 0$ the GC orbital motion affects significantly the BHB late evolution. Indeed, in this case, the $1-e_\bhb^2$ term sharply decreases down to $\sim 60\%$ of its initial value for high GCs eccentricities. This effect is enhanced for moderate values of $e$, leading such term to $\sim 30\%$ of its initial value. This implies that the BHB eccentricity grows over time, thus causing a periodic decrease of the BHB pericentral distance.
Our findings are again in agreement with \cite{quin96} results, who showed that the eccentricity growth rate is maximized for mild eccentricity values, and decreases down to zero for highly eccentric orbits.

The evident increase of $e_\bhb$ seems related to both the GC and BHB initial eccentricities. At increasing $e_\bhb$ values correspond a larger decrease of the $(1-e_\bhb^2)$ factor. For a nearly radial BHB, a high GC eccentricity leads to a less evident increase of $e_\bhb$, as evidenced comparing top and bottom right panels in Figure \ref{f2}.
The spikes that are occasionally observed in the $a_\bhb$ and $e_\bhb$ evolution mark the time at which the GC core, or what remains of it, passes at pericentre, perturbing temporarily the BHB orbital parameters. 

The initial BHB and GC orbits affect significantly also the BHB semi-major axis evolution. In models with $e_\gc = 0.5$, we found that $a_\bhb$ reduces more efficiently, being $\sim 10\%$ smaller than in models with $e_\gc = 0.9$ calculated at a time $t = 500 T_\bhb$.

A more efficient shrinking arising from models with a lower $e_\gc$ is expected, due to our choice of keeping the GC pericentre constant. This choice implies that in models with $e_\gc=0.9$ the GC moves on a less bound orbit, thus the energy that can be extracted from the BHB in three-body scatterings is smaller and the BHB evolution is less evident.

Analogously, the difference in the $a_\bhb$ at varying $e_\bhb$ are due to our choice to keep, in our models, the initial BHBs apocentre $r_{a,\bhb} = a_\bhb(1-e_\bhb) = 1$ pc.
Therefore increasing $e_\bhb$ means increasing its binding energy $E$. As the BHB energy change per GC passage is approximately constant, the ratio $\Delta E/E$ becomes smaller and the BHB hardening appears less efficient than in the case of its circular orbit.   

Note that the initially sharp decrease observed in all the cases marks the moment in which the GC impacts the BHB. This time-scale separates two different evolutionary regimes for the BHB: an initial ``violent'' phase, which cause an abrupt decrease of $a_\bhb$ and increase of $e_\bhb$ due to the GC flyby, and a second ``secular'' phase, driven by the interaction with GC stars.

\cite{goicovic17} reported a similar behaviour in the case in which a gas cloud with mass $0.01 M_\bhb$ impacts a BHB moving in a circular orbit. The similarity between hydrodynamical and direct $N$-body simulations suggests that the engine driving mostly the BHB evolution is the dynamics. However, gaseous processes, like accretion of matter onto the BHs, can speed up the BHB evolution due to the larger amount of GWs radiated away, provided that the accreted gas is fuelled at a relatively large rate $1-10\Ms$ yr$^{-1}$ for $0.1-1$ Gyr \citep{goicovic17}.  

Almost in parallel with our work, \cite{goicovic18} provided a new set of hydrodynamical simulations investigating the cumulative effects of multiple clouds impacting the BHB. 
Similarly to our results, they found that the clouds arrival marks an evident change in the BHB evolution, while the long-term evolution is mostly dominated by the BHB mass increase due to matter accretion.

\subsubsection{The effect of the BHB total mass and mass ratio}

When the GC-to-BHB mass ratio is $0.01$, the hardening efficiency is much smaller, as expected from the smaller value of $M_\gc/M_\bhb$. 
In this case, the BHB evolution is poorly affected by one single GC, but can be significantly altered if the GC mass is larger, or if many GCs reach the BHB before getting disrupted by tidal forces.

\begin{figure}
\includegraphics[width=8cm]{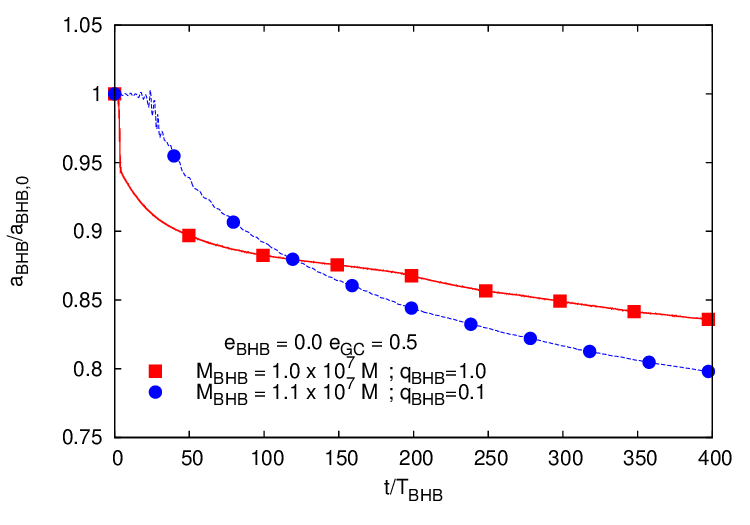}
\caption{Time evolution of the BHB semi-major axis in two models with different BHB mass ratio.}
\label{f4}
\end{figure}

As stated above, in all our models the BHB is assumed to be an equal mass binary, but one, where the BHB mass ratio is set to $0.1$. In the latter run, the BHB total mass is $1.1\times 10^7 \Ms$, i.e. the secondary SMBH has the same mass of the infalling GC, and we assumed $e_\bhb=0$ and $e_\gc=0.5$. Figure \ref{f4} shows how $a_\bhb$ varies in time in this case, compared to the evolution of the corresponding model with $10^7 \Ms$. 

A smaller mass ratio is characterised by a larger hardening rate, while the evident shift between the two curves in Figure \ref{f4} is related to the time normalization factor, which we chose as the BHB period, slightly different in the two runs.

We note that an unequal mass binary has a smaller binding
energy for a given $a_\bhb$ value, hence an unequal mass binary
shrinks more than an equal mass binary at fixed BHB mass. 
Moreover, an unequal mass binary evolves slower since the evolution
is mostly driven by the interactions between stars and the secondary, which
has a smaller cross section compared to the equal mass case \citep{sesana08}.

An interesting test case for unequal mass binaries can be found in our neighbourhoods. Indeed, the Milky Way is expected to merge with Andromeda within approximately 4 Gyr \citep{Cox07}. The M31 SMBH and Sgr A* will rapidly spiral into the merger remnant, eventually forming a binary system characterized by a very small mass ratio $M_{\rm SgrA*}/M_{\rm M31} = 0.05$. 
In order to further explore the response to a massive perturber flyby by such a low mass ratio BHB, we run a model tailored on the Milky Way and M31 SMBHs, as discussed in Section \ref{milkomedasec}.

\subsubsection{The role of GC-BHB mutual orientation}

In order to investigate the role played by the GC-BHB mutual configuration, we ran 3 models assuming $i=0^\circ$, thus an intiially prograde system. Note that we are assuming model No. 4 as reference case, thus being $M_\bhb = 10^7\Ms$, $a_\bhb = 0.9$, and $e_\gc = 0.5$.

\begin{figure}
\includegraphics[width=8cm]{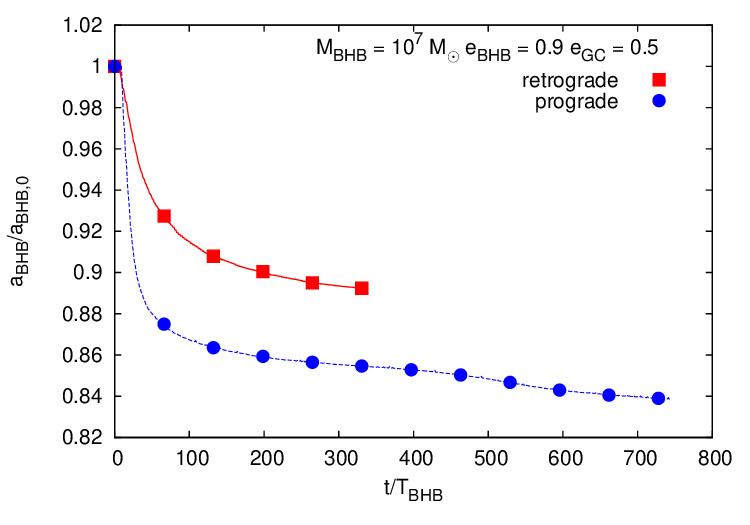}
\includegraphics[width=8cm]{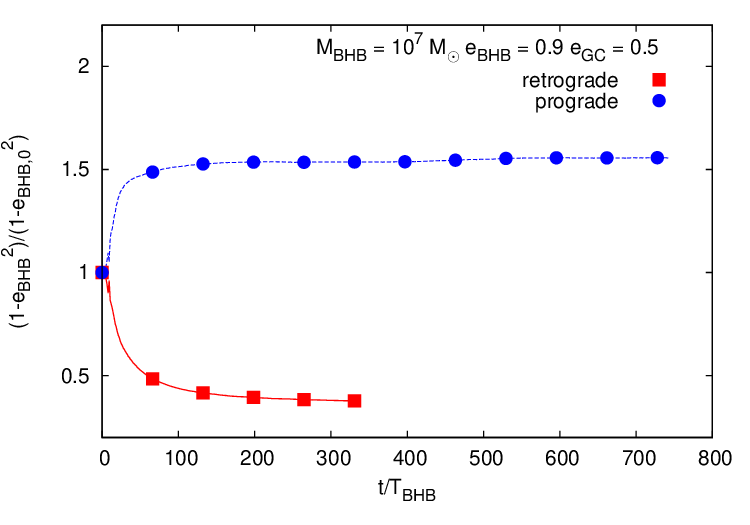}
\caption{Time evolution of $a_\bhb$ (top panel) and $1-e_\bhb^2$ (bottom panel) in a co- (model 6) and counter-rotating (model 7) configuration.}
\label{f5}
\end{figure}

Figure \ref{f5} compares the BHB semimajor axis and eccentricity evolution in co- and counter-rotating systems. The comparison outlines two important information:
\begin{itemize}
\item co-rotating configurations allow a more efficient decrease of $a_\bhb$;
\item co-rotating configurations lead the BHB orbit to circularize.
\end{itemize}
In all the co-rotating configurations that we modelled, we found the same trend, independently on the BHB mass or the GC eccentricity.

In the reference case shown in Figure \ref{f5}, we found that prograde orbits affect mostly the evolution of $a_\bhb$, which reduces by a factor $50\%$ larger than in the corresponding retrograde case.
Hence, our results suggest that the BHB hardening depends on the BHB-GC orbital configuration, being more efficient in the prograde case.
The orbital configuration has significantly different effects on the BHB eccentricity evolution. Indeed, while prograde orbits tends to decrease $e_\bhb$, driving the BHB toward circularization, retrograde configurations boost the BHB eccentricity increase up to $e_\bhb = 0.97$. 
As a consequence, the product $a_\bhb^4(1-e_\bhb^2)^{7/2}$ decrease much faster in counter-rotating systems than in co-rotating, thus reducing more efficiently the GW time-scale. In the example shown above, the final $t_\gw$ in the retrograde model is $\sim 37$ times smaller than in the prograde case.
Hence, counter-rotating configurations seems to enhance the BHB contraction and, possibly, their coalescence.

The BHB response to prograde or retrograde perturbers is different if the impacting system is a gaseous cloud. Indeed, in this case the accretion of retrograde gas erases the BHB angular momentum, making its shrinkage more efficient than in the prograde case \citep{goicovic18}.

The BHB eccentricity increase driven by retrograde orbits is related to the combined action of the BHB gravitational field and of the outer star. Indeed, as each star in the GC approaches the BHB, it is subjected to a periodic oscillation of the gravitational field in which it is travelling, which leads to a torque exerted perpendicularly to the star orbital plane  \citep{merritt09}. In the case of an unequal mass BHB, this effect causes a change in the secondary BHB angular momentum and, in turn, a variation of the orbital eccentricity. As shown by \cite{sesana11} through three-body scattering experiments and theoretical arguments, the resulting eccentricity variation of the BHB eccentricity on pc-scales has different amplitude and sign in co-rotating and counter-rotating configurations, leading to an increase in the counter-rotating case.
Recently, \cite{bockelmann15} found the same behaviour modelling the evolution of an BHB embedded in a rotating galaxy disc, on kpc scales \citep[but see also][]{rasskazov17}.

Our models demonstrate that this effect still works efficiently on pc scales, driving, as we will show in the next section, the BHB to coalescence in several cases. 

\begin{figure}
\includegraphics[width=8cm]{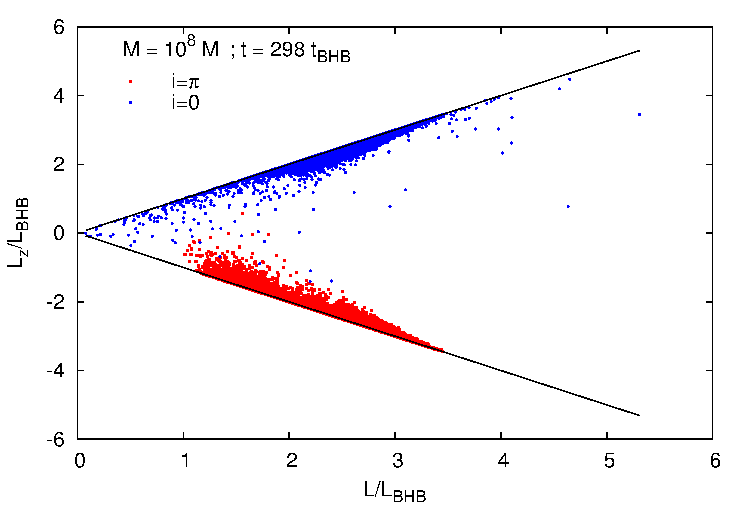}
\caption{Angular momentum of GC stars after its disruption. The BHB mass in this case is $10^8\Ms$, while the eccentricities are $e_\bhb = e_\gc = 0.5$.}
\label{f6}
\end{figure}

The distribution of stars angular momentum in the phase space is well defined after the GC passage at pericentre, and strongly depends on its initial inclination, as shown in Figure \ref{f6}. 
Due to the co-planarity assumption, initially both the BHB and GC angular momentum vector lies on the z-axis. 
After the GC flyby, some stars gain a small component on the x and y axes, while their distribution in the $L_z-L$ plane changes significantly depending on the initial GC inclination. Note that the cluster stars in the counter rotating system distribute in a more concentrated region of the phase space, with almost all the stars having $1<L/L_\bhb<3.5$. Also, the angular momentum diffusion seems more efficient, with stars gaining non-zero x and y components. At low inclinations the stars angular momentum spans a wider range, with values $0<L/L_\bhb<6$, but its diffusion is less evident. 

In order to provide further insights on the role played by the GC orbital inclination and its pericentre, we ran a supplementary set of three simulations, using model No. 4 as a reference ($M_\bhb=10^7\Ms$, $e_\bhb = 0.5$ and $e_\gc = 0.9$).
We label these further models with numbers 18, 19 and 20, as shown in Table \ref{tab1}.

\begin{figure}
\centering
\includegraphics[width=8cm]{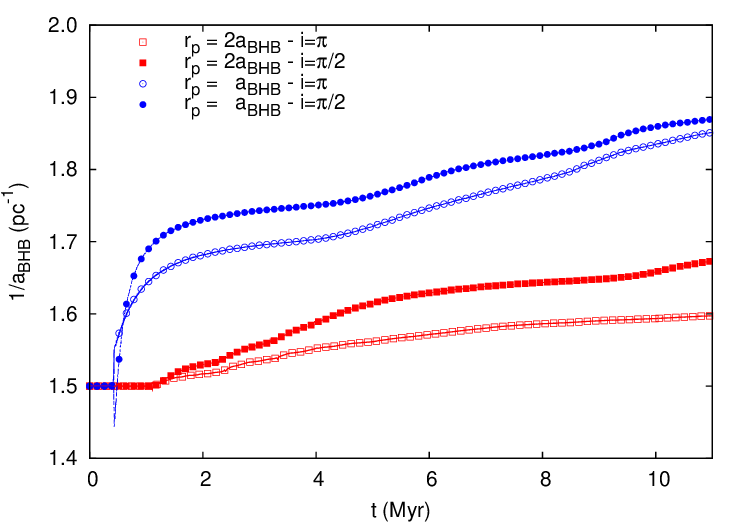}\\
\includegraphics[width=8cm]{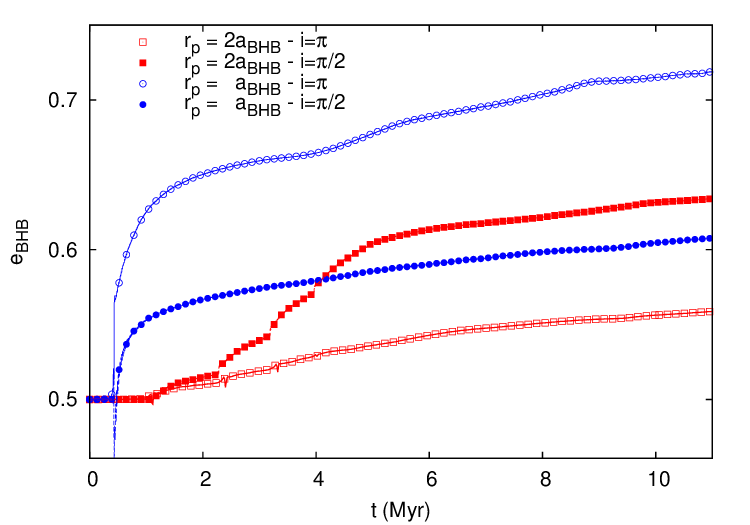}
\caption{Time evolution of the BHB inverse semi-major axis (top panel) and eccentricity (bottom panel) at varying the
GC pericentral distance and the GC orbital inclination.}
\label{mutua}
\end{figure}

In model No. 18, we assumed that the GC orbital plane is perpendicular to the BHB orbit ($i = 90^\circ$), keeping fixed all the other parameters. In model No. 19, instead, we assumed that the GC moves on a retrograde orbit characterized by a pericentre $r_{\rm p,GC} = a_\bhb$, thus half of the value assumed for the reference model. Similarly, we assumed $r_{\rm p,GC} = a_\bhb$ also for Model No. 20, with the further constraint that the BHB and the GC orbital planes are perpendicularly oriented. 
Figure \ref{mutua} shows how the semi-major axis and eccentricity evolve for these supplementary models and for the corresponding reference.

Interestingly, it is evident that when the GC moves on a perpendicular orbit flyby stars extract energy
from the BHB with a higher efficiency, enhancing the BHB shrinkage. This enhanced hardening is less evident when the GC pericentre becomes comparable to the BHB semi-major axis. 
The net effect on the BHB evolution is relatively small, being only $5 \%$ than the retrograde case when $r_{\rm p,GC} = 2a_\bhb$, and almost negligible ($1 \%$) when $r_{\rm p,GC} = a_\bhb$. Note that a similar difference is observed comparing retrograde and prograde orbits, as shown in Figure \ref{f5}. Hence, the hardening efficiency of a perpendicular orbit lays between the prograde and retrograde cases.
A perpendicular orbit also affects significantly the BHB eccentricity evolution, which shows an interesting dependence also on the GC pericentre.
Indeed, in the case $r_{\rm p,GC} = 2a_\bhb$, after the GC close flyby the BHB eccentricity increase is faster than in the retrograde case, reaching values above $\gtrsim 0.62$, thus implying a reduction of the GW time-scale by a factor $((1-e_{\rm ret}^2)/(1-e_{\rm per}^2))^{7/2}\sim 1.5$.
However, it appears evident that the eccentricity increase depends also on the GC pericentre. Indeed, when the GC get closer to the BHB, a retrograde orbit drives a very fast and effective eccentricity increase, leading to values above $e_\bhb > 0.7$, while the increase is much smaller in the perpendicular case, as shown in the bottom panel of Figure \ref{mutua}.

The results discussed above highlight the importance of the orbital parameters in
determining the GC-BHB interaction and long-term evolution.

\subsubsection{The net effect on the BHB orbital evolution}

\begin{figure}
\includegraphics[width=8cm]{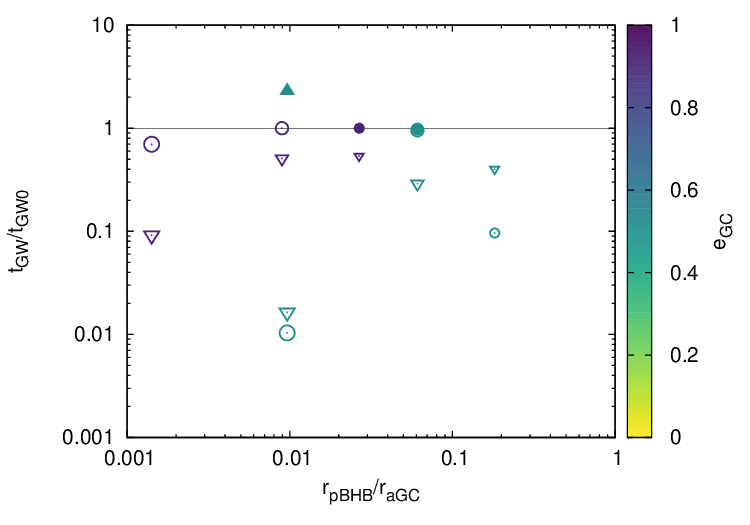}
\caption{GW time, normalized to its initial value, as a function of the ratio between the BHB pericentre and the GC apocentre. The color-coded map identifies the GC eccentricity. Smaller points corresponds to smaller values of $e_\bhb$. Filled points label co-rotating models, while open points refer to counter-rotating systems.Triangles refer to $M_\bhb = 10^7\Ms$, while circles refer to $10^8\Ms$. Two out of the three co-rotating cases overlap the counter-rotating counterparts.}
\label{f7}
\end{figure}

As discussed above, the GC debris accumulate around the BHB causing in several cases its hardening and the eccentricity increase, with an efficiency that depends on the set of initial conditions assumed. In order to determine the significance of such an effect on the BHB coalescence, we plot in Figure \ref{f7} the ratio between the initial and final value of $t_\gw$, as evaluated through equation \ref{peters}, as a function of the BHB pericentre, normalized to the GC apocentre. All the points are calculated at an integration time $10^3$ times the BHB initial orbital period.
We note that in co-rotating systems the BHB evolution is weakly affected by the GC, even softening in one of the tested cases. Regarding counter-rotating configuration, the maximum efficiency in the BHB hardening is achieved for GCs moving on a mildly eccentric orbit, $e_\bhb \sim 0.5$, and the BHB eccentricity is high. In this case, the GW time-scale reduces by a factor 100, quite independently on the BHB mass, which for the $M_\bhb=10^8\Ms$ model translates into $t_\gw = 5.4$ Gyr, smaller than a Hubble time. 

Our results points out that the GC impact can potentially enhance the BHB hardening rate and facilitate its merger, especially in the case in which the binary has an initially high eccentricity.

Recently, \cite{bortolas17} conducted a similar work, modelling the evolution of a young cluster with mass $8\times 10^4\Ms$ impacting the BHB on a purely radial or highly eccentric orbit ($e>0.75$) moving in a co-planar or perpendicular configuration. Their cluster model is initially characterized by a broad mass spectrum and a phase-space distribution drawn accordingly to a \cite{King} model with $W_0=5$ and core radius $0.4$ pc. Their BHB is an equal mass binary with total mass $10^6\Ms$ moving on a circular orbit. We note that the BHB-to-GC mass ratio in this case is $0.04$. 
As we did, the authors take into account the effect of an external potential, modelled according to a Dehnen potential tailored to the Milky Way galactic halo. 
In their simulations, \cite{bortolas17} found that only purely radial orbits can bring the BHB in the GW emission regime, while clusters moving on eccentric orbits only marginally affect the BHB evolution.

We stress here that our 16 models allow investigating a much larger portion of the phase-space.
Our investigated BHB-to-GC mass ratio ranges in between $10^{-2}-10^{-1}$, thus including the \cite{bortolas17} models, while our galaxy and BHB assumptions are tailored to a massive elliptical galaxy, rather than the Milky Way halo. Moreover, we briefly discuss the importance of the BHB mass ratio, modelling the future evolution of the massive BHB that will form in consequence of the Milky Way and Andromeda galaxy merging.
Some of our results overlap with \cite{bortolas17} paper, thus providing a further confirmation of the agreement between the different numerical methods used.
We note that our results based on circular BHB and $M_\bhb/M_\gc=0.01$ are in good agreement with their results, as we did not find any significant shrinkage of a circular BHB interacting with a GC moving on an eccentric orbit but, on the other hand, we found that this effect works much more efficiently at higher BHB-to-GC mass ratio and non-zero BHB eccentricities.

The condition of having $e_\bhb>0$ is supported by the fact that the two SMBHs likely bound together in consequence of a galaxy merger, thus the SMBHs mutual orientation when they come sufficiently close to form a binary system are inherited, at least in part, from the original galaxy merger configuration. 
Our results suggest that if the BHB undergoes a strong scattering with a massive perturber before it circularize, the effect can be sufficiently strong to induce the BHB to coalesce within a few Gyr.

In the next section we focus on the dynamical feedback that the BHB impinges on the GC stellar debris.

\subsection{The production of high-velocity stars}
\label{hyperS}
We notice that in all the models investigated the GC-SMBH flyby drives the ejection of several stars with velocities varying between $10^2-10^3$ km s$^{-1}$, compatible with observed high and hyper-velocity stars (HVSs). 
A HVSs origin driven by GC impacting single or double SMBHs, a channel already discussed in several papers \citep{ASCDS16,ASCD17,cd15,fragione16,fragione17,bortolas17}, would be easily observable in term of HVSs location and velocity distribution \citep{ASCD17c,fragione17}. 
The role played by this channel in creating HVSs can be unveiled by the GAIA\footnote{\url{http://sci.esa.int/gaia/}} Data Release 2 \citep{marchetti17}. 

If a fraction of HVSs are formed through GC-BHs interactions, their angular momentum and location in the galaxy may contain some information on their host cluster \citep{fragione16,ASCD17c,fragione17}. However, the observational signatures of HVSs produced through this channel might be similar to those observed for standard star emission by BHBs by gravitational slingshot \citep[see][]{sesana06}.

\begin{figure}
\centering
\includegraphics[width=8cm]{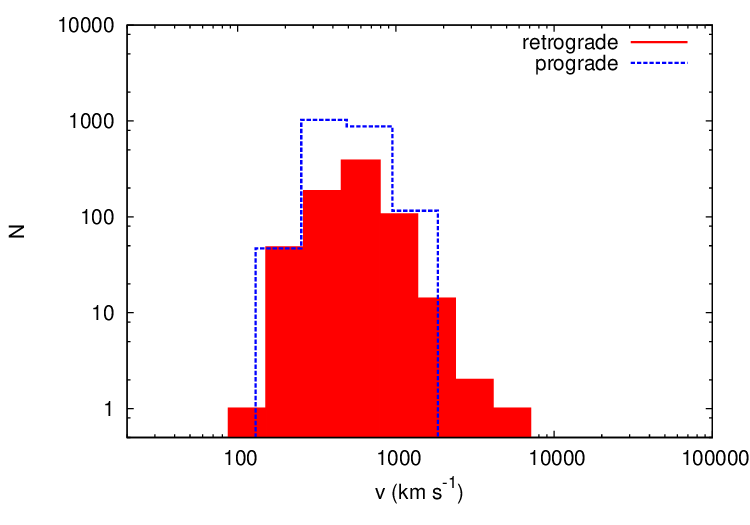}
\caption{Velocity distribution of escaper stars in case $M=10^7\Ms$, $e_\bhb = 0.9$ and $e_\gc = $ for the retrograde (red filled steps) and prograde (blue dashed steps) configuration. Stars velocities are calculated locally, i.e. in the star position, after $t = 2000 T_\bhb$, provided that the star position is outside 1 kpc from the galaxy.}
\label{esc_retro}
\end{figure}

Although a detailed study of the relations between HVSs and GC-BHB is out of the scopes of this paper, we notice here that in our models $\sim 1-3\%$ of the GC stars are ejected in the galaxy outskirts and would appear like HVSs, observationally. Hence, a large number of HVSs moving in a galactic halo can be the signature of a massive BHB harboured in the galactic centre.
Most notably, as shown in Figure \ref{esc_retro}, we found an evident difference between retrograde and prograde models in the number and velocity distribution of escaper stars.
Indeed, stars ejected in the prograde case are characterized by a narrower velocity distribution, with maximal velocities up to $1800$ km s$^{-1}$, while in the retrograde case the tail of the distribution extends beyond $2000$ km s$^{-1}$. Ejected stars in the prograde model outnumber those in the retrograde case, being $N_{\rm pro} = 2.8 N_{\rm retro}$. 

This is likely due to the fact that stars moving on co-rotating orbits are easier to eject due to their smaller relative velocities and larger gravitational focusing effect \citep{iwasawa11}. The BHB angular momentum is transferred much more efficiently to stars moving on retrograde orbits, being such transfer the main cause of the BHB eccentricity grow in the retrograde case \citep{iwasawa11}.

\subsection{The possible formation of a NSC around a massive BHB}
\label{ncform}
\begin{figure}
\includegraphics[width=8cm]{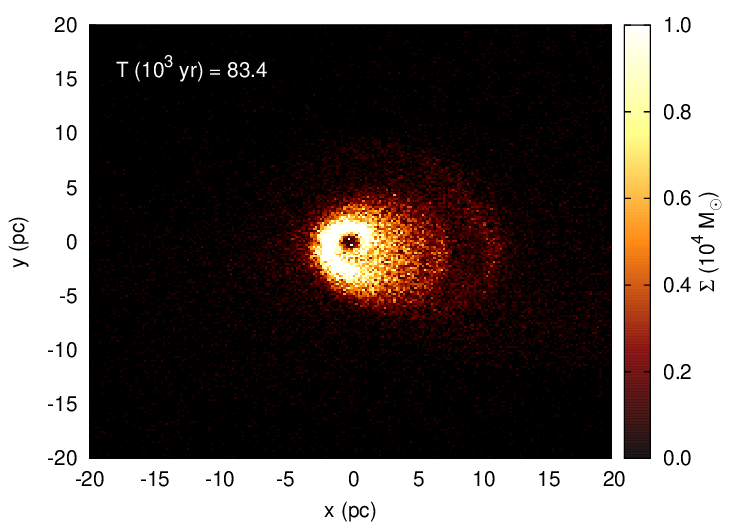}\\
\includegraphics[width=8cm]{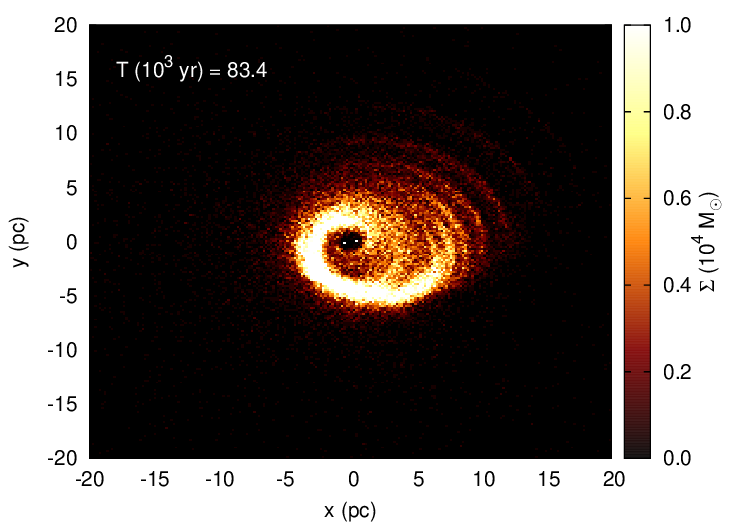}\\
\caption{Snapshot of the GC stars distribution after the flyby. The BHB mass is $M=10^7 \Ms$ (top panel) and $M=10^8 \Ms$ (bottom panel), while in both cases $e_\bhb=e_\gc=0.5$. }
\label{f8}
\end{figure}

After the GC flyby, some stars bind to the BHB and contribute to its hardening, while some other are pushed on larger orbits. 
In all the models, the GC debris assume a disc structure, characterized by several ``rings'' of stars which are accelerated by the BHB (see Figure \ref{f8}). The life-time of such a structure depends on the BHB mass and eccentricity, being efficiently disrupted at higher values of $M_\bhb$.

It is worth noting that the ring-like structures are much more evident at smaller $M_\gc / M_\bhb$ values.
Similar features are observed also in the case of a gaseous impactor \citep[see for instance][]{goicovic16} assuming a cloud-to-BHB mass ratio $0.01$, thus indicating that they are likely due to the efficient transfer of energy and angular momentum from the BHB to the impactor debris.

An efficient deposit of mass from infalling dense clusters is at the basis of the NSC formation in the framework of the so-called ``dry-merger'' scenario \citep{Trem76,Dolc93}. However, its validity has not been investigated extensively in galaxy models hosting a massive BHB. 
Using numerical simulations, \cite{bekki10} showed the evolution of a massive BHB can efficiently heat a NSC after a galaxy merger, leading to its erosion and disruption.
However, it is not clear whether the NSC can re-generate due to the infall of a star clusters after the BHB formation.

Our current knowledge of star clusters formation and evolution suggests that the total mass in GCs is roughly $1\%$ of the total hosting galaxy mass \citep{harris10}. 
Therefore, a galaxy with typical mass of $10^{11}\Ms$ is expected to host $\sim 10^{4}$ clusters, assuming an average GC mass of $\sim 10^5\Ms$.
The time-scale over which GC orbitally segregate to the galactic centre is given by \citep{ASCD14a,ASCD15He}:

\begin{align}
\label{tdfeq}
t_{\rm df} =& 0.3 {\rm Myr} \left(\frac{r_g}{1 {\rm kpc}}\right)^{3/2} \times \\ 
\nonumber
 & \times \left(\frac{M_g}{10^{11}\Ms}\right)^{-1/2}
g(e_\gc,\gamma)\left(\frac{M_{\rm GC}}{M_g}\right)^{-0.67}\left(\frac{r_{\rm GC}}{r_g}\right)^{1.74},
\end{align}
where $g(e,\gamma)$ is a function of the GC orbital eccentricity $e_\gc$ and the galaxy density slope $\gamma$.

\begin{figure}
\centering
\includegraphics[width=8cm]{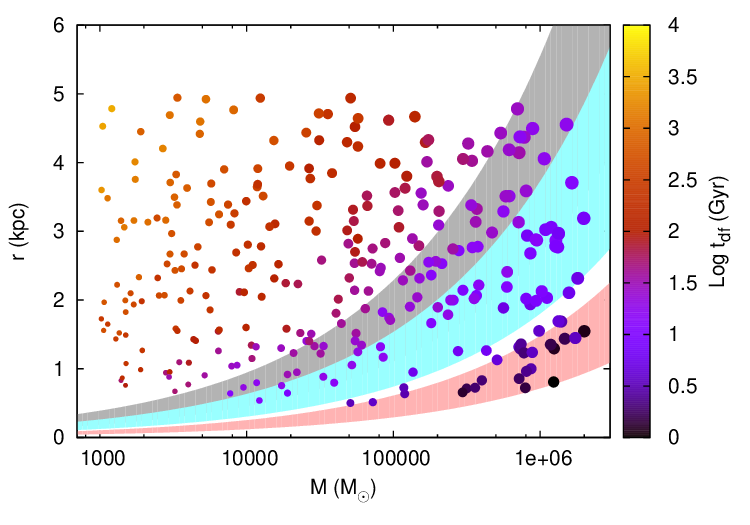}
\caption{Mass (x-axis) and apocentre (y-axis) for a sample of 300 GCs in our galaxy model. The color coded map represent the dynamical friction time-scale according to \citep{ASCD14a}. The different sizes of the dots refers to different GCs tidal radii. The shaded areas identify regions at constant $\tau_{\rm df} = 1$ Gyr (red area), $5$ Gyr (cyan area), and $10$ Gyr (grey area). The upper and lower area limits are defined by radial and circular orbits, respectively.}
\label{tdfmap}
\end{figure}

Under the assumption that GCs distribution in the phase space follows the galaxy distribution, it is possible to show using Equation \ref{tdfeq} that 
$\sim 16\%$ of the GCs undergo orbital decay within a Hubble time. 
For the sake of clarity, we show the orbital decay time-scale $t_{\rm df}$  in Figure \ref{tdfmap} for a subsample of 300 GCs in our galaxy model. 

Shaded areas highlight the regions in the $M-r$ plane for which $t_{\rm df} = 1-~5-~10$ Gyr. Circular orbits set the area's lower limit, while radial orbits represent the upper limit.

\begin{figure}
\includegraphics[width=8cm]{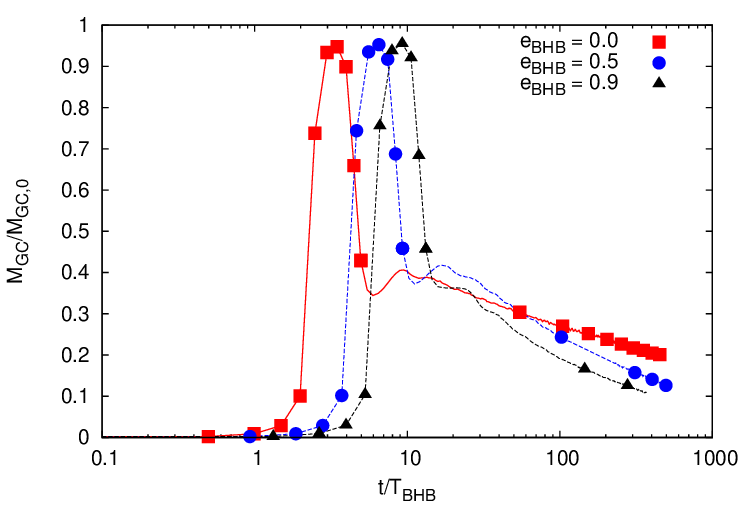}\\
\includegraphics[width=8cm]{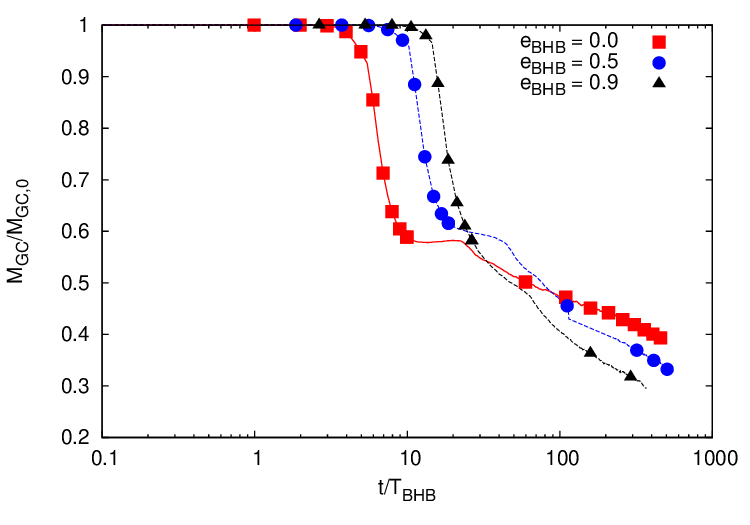}\\
\includegraphics[width=8cm]{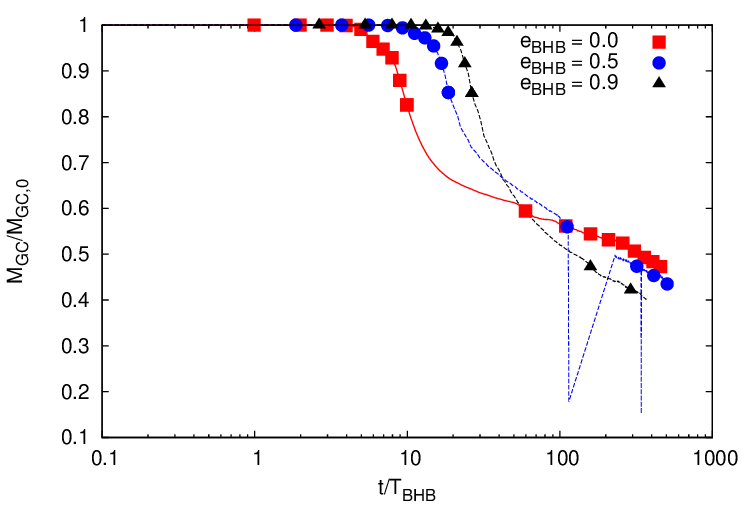}\\
\caption{GC mass, normalized to its initial value, deposited within 4 pc (top panel), 10 pc (central panel) and 20 pc (bottom panel) from the BHB as a function of time and for different values of the eccentricity. All the three models are characterized by $M_\bhb = 10^8\Ms$ and $e_\gc = 0.5$.}
\label{f9}
\end{figure}

\begin{figure}
\includegraphics[width=8cm]{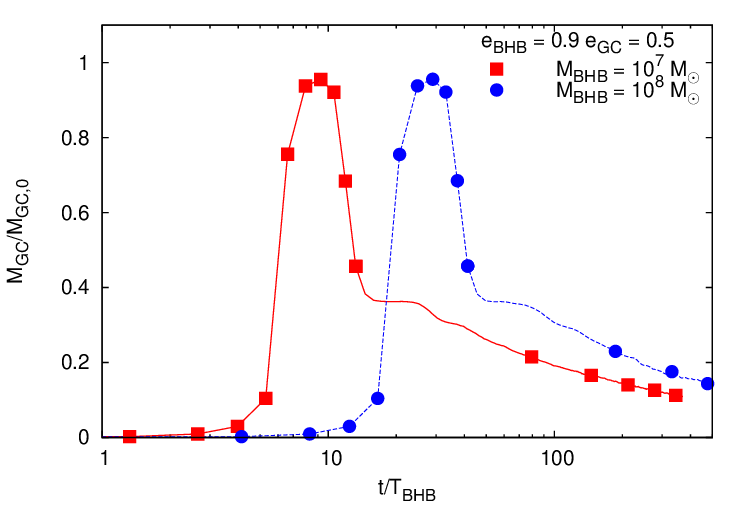}
\includegraphics[width=8cm]{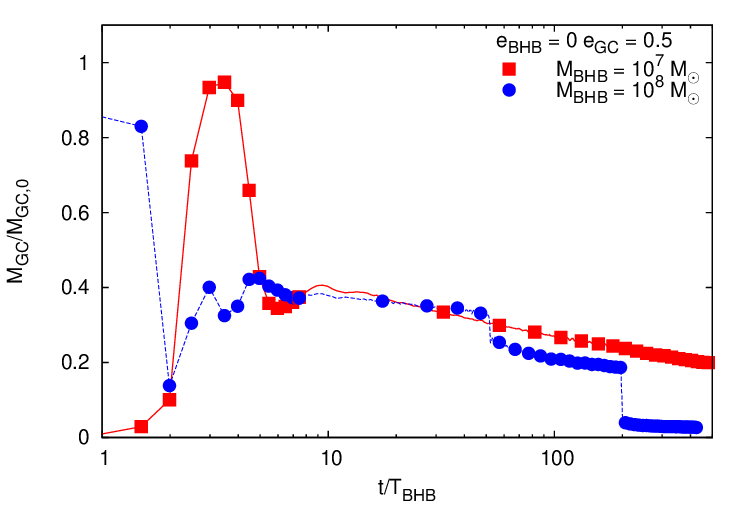}
\caption{GC Mass, normalized to its initial value, enclosed within 4 pc from the BHB in two models with $M_\bhb=10^7\Ms$ (red solid line) and $10^8\Ms$ (blue dashed line). Top panel refers to $e_\gc = 0.5$ and $e_\bhb=0.9$, while in bottom panels $e_\gc = 0.5$ and $e_\bhb=0$.}
\label{f10}
\end{figure}

A good proxy to follow the possible formation of an NSC seed into the galactic nucleus is represented by the mass accumulated within the inner few pc around the BHB. 
Figure \ref{f9} shows the time evolution of the mass deposited within 2 pc from the BHB at varying values of $e_\bhb$, for $M_\bhb = 10^8\Ms$ and $e_\gc = 0.5$.
The evident peaks present in the three cases mark the time at which the GC reaches the galactic centre. 
Since the time here is normalized to the BHB orbital period, which varies accordingly to $e_\bhb$ and the semi-major axis, the peaks are not overlapped. 
After the GC passage at pericentre, some mass continues moving on its orbit, leaving the 4 pc sphere, while nearly the $40\%$ of the cluster mass remains ``trapped'' around the BHB, being gradually depleted over time and reducing to less than $20\%$ in all the cases, as shown in the top panel of Figure \ref{f9}. This is likely due to the slingshot of stars operated by the BHB. Looking at the central and bottom panels, which show the time evolution of the GC mass enclosed within 10-20 pc from the BHB, makes evident the efficiency of the slingshot effect.
The BHB carving action maximizes at increasing $e_\bhb$ values, as we found that the GC mass orbiting the inner 10 pc 
after the GC disruption is $\sim 40\%$ if $e_\bhb = 0$, but decreases below to $30\%$ if $e_\bhb = 0.9$.
On the other hand, it is quite clear from the plots that the stars removal operated by the BHB in our simulations has not saturated, and works still efficiently on a time-scale $\sim 10^3$ times the BHB orbital time. This effect is even more efficient when $M_\bhb=10^8$, as in this case the mass accumulated within 10 pc from the BHB barely exceeds $10\%$, as shown in Figure \ref{f10}. 

\begin{figure}
\centering
\includegraphics[width=8cm]{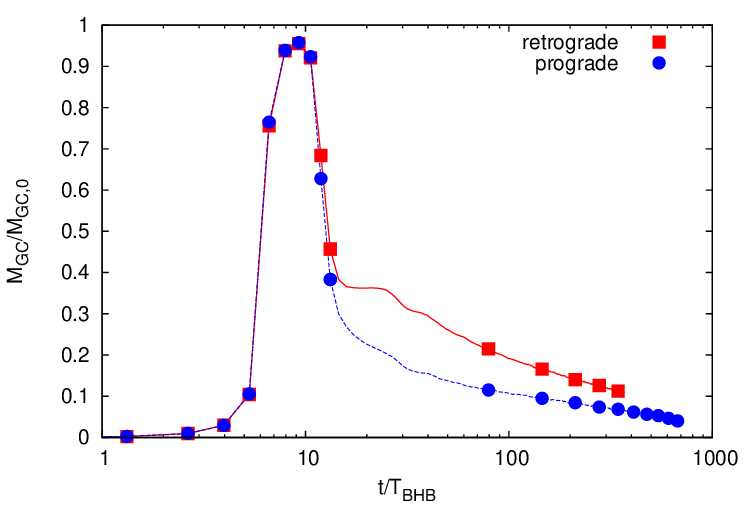}
\caption{Deposited mass in the case of $M_\bhb=10^7\Ms$, $e_\gc = 0.5$, $e_\bhb=0.9$ in an initially counter-rotating  (red straight line) and co-rotating configuration (blue dotted line).}
\label{f11}
\end{figure}

Also the relative inclination of the GC orbit matters in determining the fate and the shape of the NSC. Figure \ref{f11} shows the enclosed mass (within 4 pc) in two models initially moving in a counter- and co-rotating configuration. It is evident that co-rotating orbits seems to deplete mass from the BHB more efficiently. The main reason for a smaller deposited mass is related to the fact that co-rotating stars have larger cross-sections compared to retrograde orbits, due to the lower star-binary relative velocity, thus being characterized by a higher probability to be ejected from the galactic nucleus. The larger ejection fraction leads to a larger amount of energy extracted from the BHB, thus explaining why in the co-rotating models we found, on average, sparser nuclei hosting harder BHBs. As opposed to this, counter-rotating models will be characterized by a denser nucleus hosting, on average, a high-eccentricity BHB. Interestingly, the long term evolution of the enclosed mass seems to converge to a common trend. This is likely due to the fact that after several BHB orbits the GC stellar debris lose information of their initial orbital configuration, due to the BHB tidal field.

\begin{figure}
\includegraphics[width=8cm]{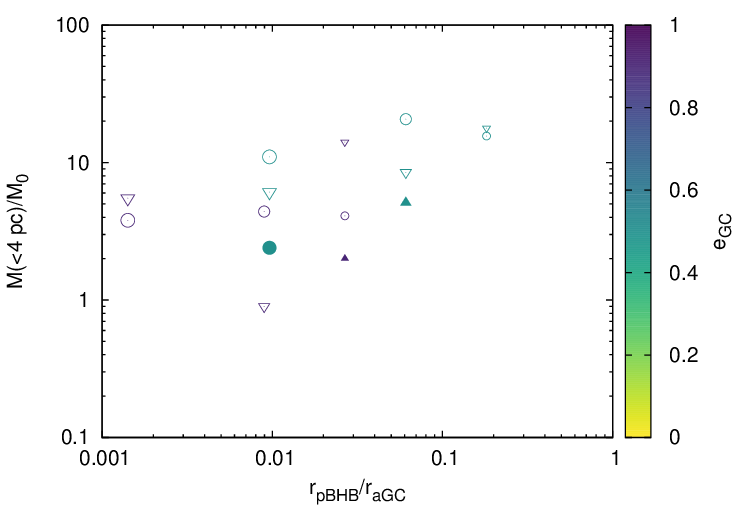}
\caption{Percentage of GC mass enclosed within 4 pc from the BHB after $10^3T_\bhb$, as a function of the ratio between the BHB pericentre and the GC apocentre. The color-coded map refers to the GC eccentricity, while at larger points correspond larger $e_\bhb$ values. Circle refers to $M_\bhb=10^8\Ms$, while triangles to $M_\bhb=10^7\Ms$. Open symbols identifies counter-rotating models, while filled symbols are dedicated to co-rotating models.}
\label{f12}
\end{figure}

The percentage of GC mass enclosed within 4 pc from the BHB is shown in Figure \ref{f12} for all the models investigated. From our data, it seems that the deposited mass increases at increasing values of the ratio between the BHB pericentre and the GC apocentre.
In all the cases, the mass deposited in the BHB surroundings never exceeds the $20\%$ of the total cluster mass. This numbers are in agreement with earlier numerical and theoretical modelling \citep{antonini13,ASCD14a,ASCDS16,ASCD17} of NSC formation around a single SMBH heavier than $10^8\Ms$.
The absence of massive NSCs in bright galaxies can be connected to the high tidal disruption efficiency operated by the compact object sit in the centre of the galaxy, either a single or double SMBH with mass $\geq 10^8\Ms$.

\subsubsection{NSC detectability}
As we have shown in Section \ref{ncform}, a substantial fraction of the GC mass is dispersed around the BHB after the scattering.
Therefore, it is quite difficult to determine whether a NSC can emerge from the galactic background if several GCs infall repeatedly onto the galactic centre.

A reliable investigation on NSC formation would require 
modelling multiple clusters and galaxy stars, making the problem largely computational demanding.
In order to explore whether a larger set of simulation is worth of further investigation, we proceeded in two ways: i) we run a new simulation comprised of two infalling clusters, ii) we developed a technique to create mock density maps in the assumption of {\it homologous} infall. 

To run the new simulation, we kept the last snapshot of model 3 ($M_\bhb = 10^7\Ms$, $e_\bhb=e_\gc=0.5$) and put on a retrograde orbit a second cluster moving on a $e_\gc = 0.9$. For comparison with other models, we kept fixed the GC pericentre at $2$ pc. 

\begin{figure}
\centering
\includegraphics[width=8cm]{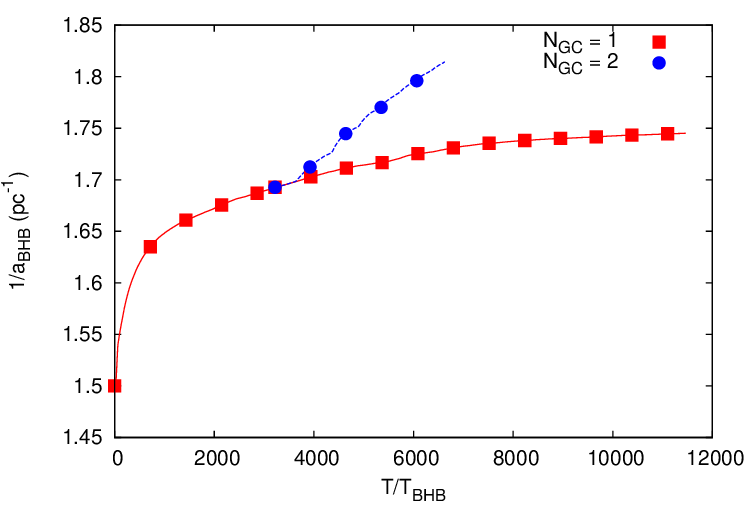}\\
\includegraphics[width=8cm]{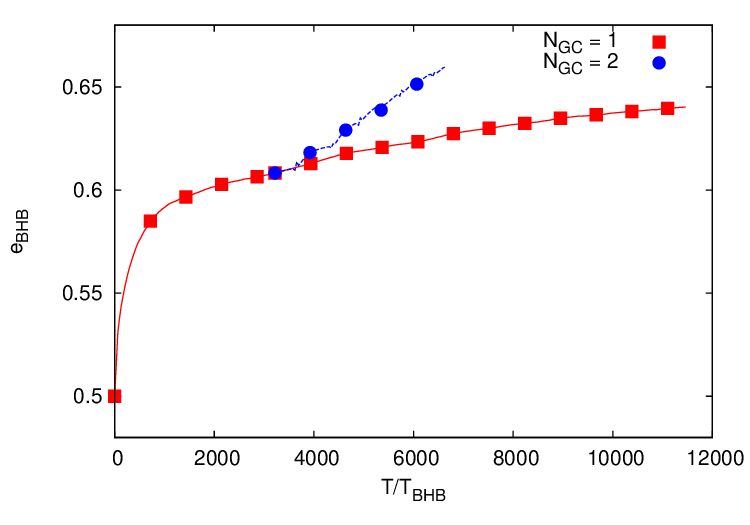}
\caption{Time evolution of the BHB inverse semi-major axis (top panel) and eccentricity (bottom panel) assuming only one infalling GC (red straight line) or two (blue dashed line).}
\label{duplex}
\end{figure}

The second GC has a fundamental impact on the BHB orbital evolution, as shown in Figure \ref{duplex}. The semi-major axis and eccentricities increases faster, compared to the case in which only one cluster is considered. The second GC undergoes several passages at pericentre before being disrupted, leaving a clear signature in the BHB eccentricity evolution. 

\begin{figure*}
\centering
\includegraphics[width=8cm]{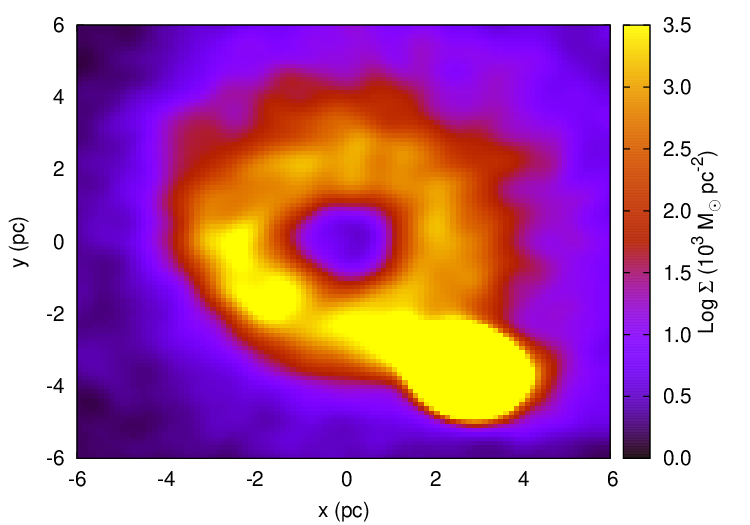}
\includegraphics[width=8cm]{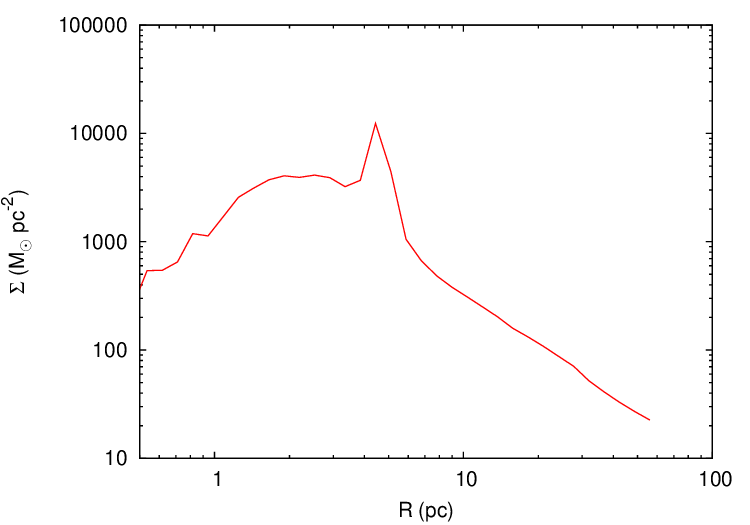}\\
\includegraphics[width=8cm]{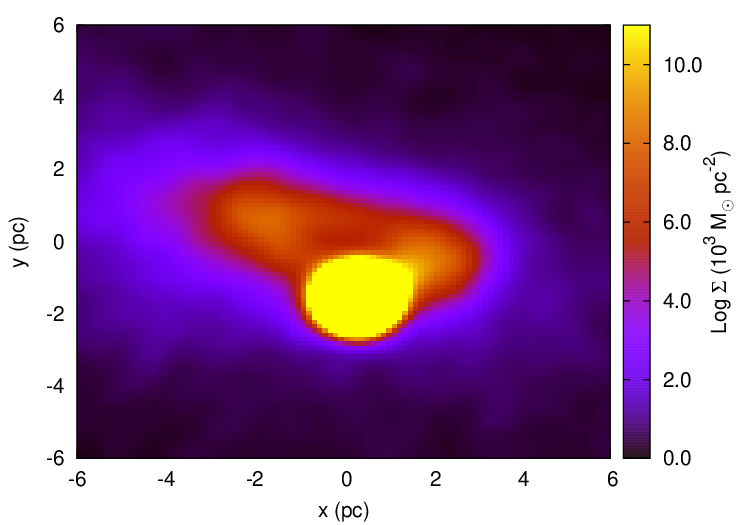}
\includegraphics[width=8cm]{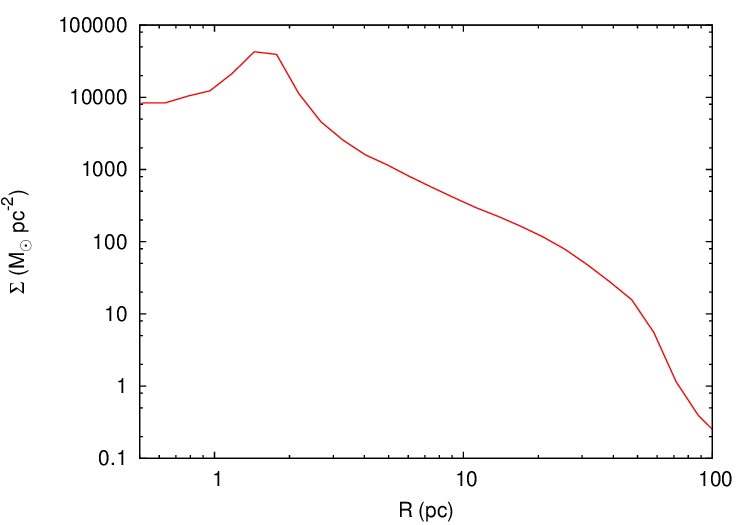}
\caption{Left panels: surface density map for the model 21 ($M_\bhb=10^7\Ms$, $e_\bhb=0.5$), characterized by two GCs having $e_\gc=0.5$ plus $e_\gc=0.9$ GCs orbits. Top panel shows a face-on view on the GC orbital plane, while in bottom panel the line-of-sight is rotated. Right panels: corresponding surface density profile.}
\label{smap}
\end{figure*}

The surface density map of the nucleus structure after $6500$ BHB orbit is shown in Figure \ref{smap}. The debris are mixed quite efficiently in the galaxy centre, making indistinguishable their previous evolution. 
The corresponding surface density profile $\Sigma(R)$ allows easily to determine whether a NSC emerges from the galactic background. 
The evident drop at distances $r<1.5$ pc evidences that the GC disruption leads to the formation of a disc structure, rather than a well defined nucleus.

Clearly, such an effect is mostly due to the line-of-sight. The left panels in Figure \ref{smap} shows how the galactic centre would look-like at varying the angle of view. In the case of a nearly face-on observation, and that the two infalling GCs move on the same plane, the galactic centre would exhibit an evident disc configuration, with the central hole extending roughly up to the BHB separation. The corresponding surface density profile increases steeply within $1$ pc, while outside such length scale decreases smoothly following the galaxy background. On the other hand, an edge on view would ``mask'' the disc structure, leading to a surface density that increases toward the galactic centre. 
In both the cases, it seems evident the presence, at $\sim 10$ pc, a transition from the inner density profile, dominated by the GC debris, and the outer, dominated by the galactic background. The surface map allows to spot the possible presence of the GCs core remnants, although these features are expected to dissolve on several GC crossing times $O(\sim 100{\rm Myr})$. 
The bright spot identifies the second GC core, which is still surviving the BHB tidal forces. 
The GC core marks the transition between the two regions depicted above, determining the clear edge at $\sim 2$ pc evident in the bottom right panel of Figure \ref{smap}.

\begin{figure}
\centering
\includegraphics[width=8cm]{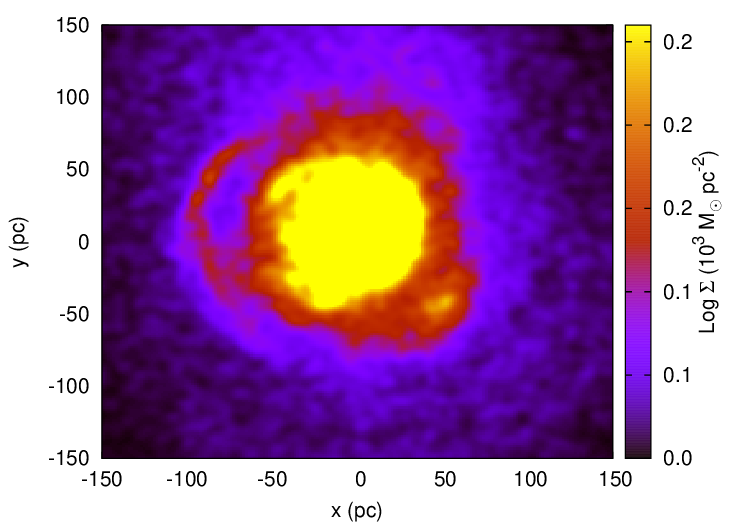}
\caption{Surface density map for model 21 (BHB + 2 GCs). A clear arch is visible at $\sim 100$ pc from the galactic centre ($x=-100;~y=-50,50$), while a clump formed in the bottom right side of the bulge ($x=50;~y\simeq -50$).  }
\label{largeR}
\end{figure}

As discussed in the previous section, a significant fraction of the GCs mass is ejected in the galactic bulge, or even outside the galaxy. The GCs debris dispersed on length-scales $\sim 100-200$ pc show extended arches whose properties are clearly related to the parent GC orbit, as shown in Figure \ref{largeR}.

The global density profile is expected to increase at increasing GCs number, possibly affecting the morphology of the galactic nucleus structure. 

\begin{figure*}
\includegraphics[width=8cm]{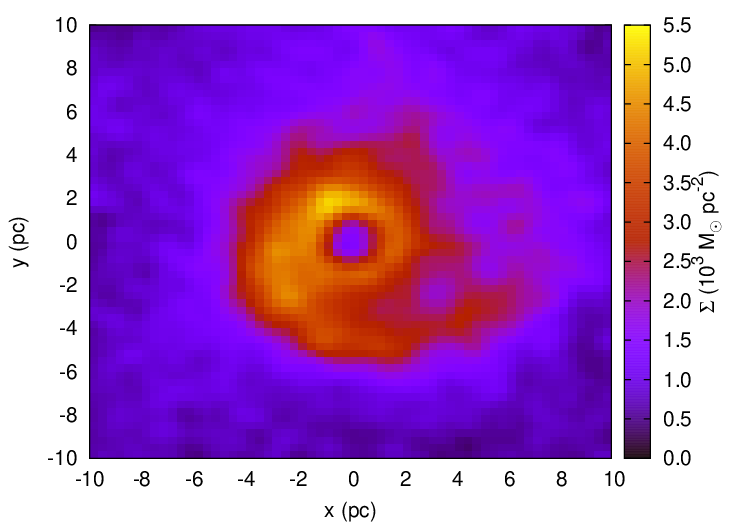}
\includegraphics[width=8cm]{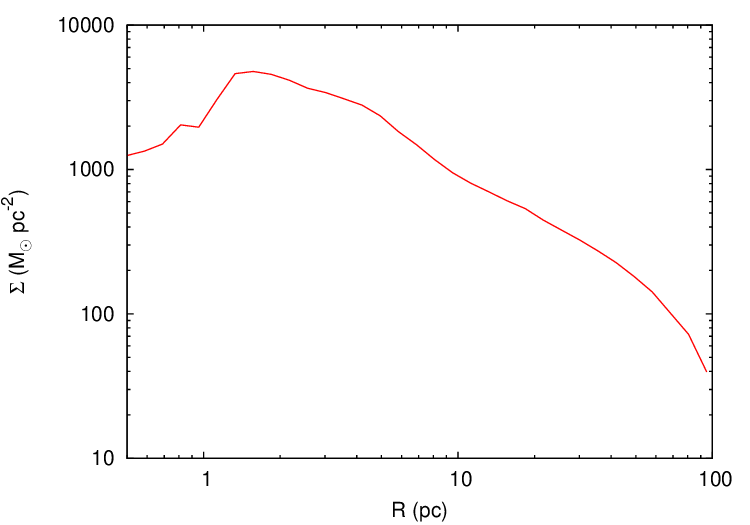}\\
\includegraphics[width=8cm]{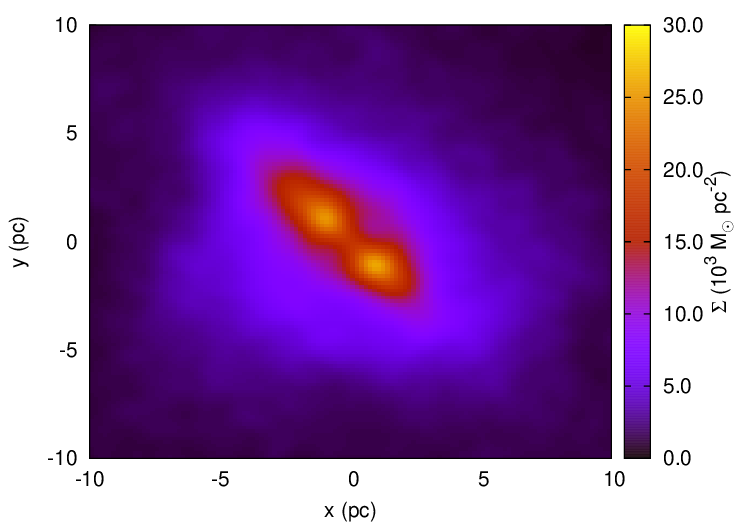}
\includegraphics[width=8cm]{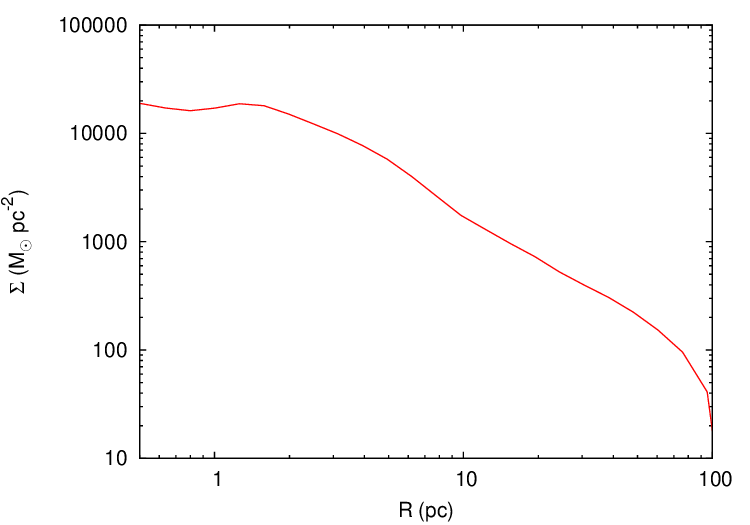}\\
\includegraphics[width=8cm]{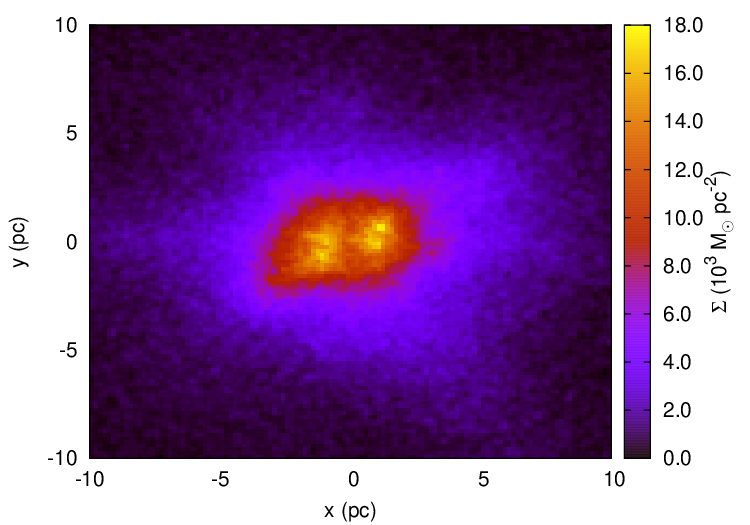}
\includegraphics[width=8cm]{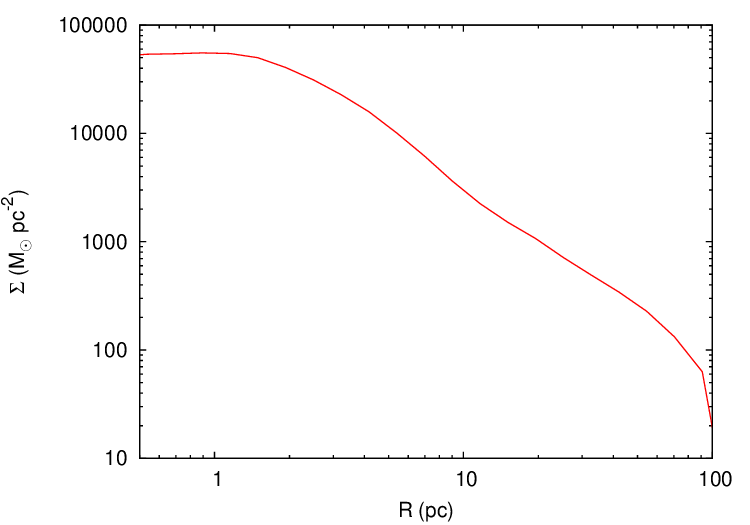}
\caption{As in Figure \ref{smap}, but assuming that the number of homologous GC infall is 2 (top panels), 4 (top panels) and 6 (bottom panels).}
\label{N6rot}
\end{figure*}

Following \cite{ASCDS16}, we used our simulations to determine whether the GC debris may leave observable signatures in the galactic nucleus. We focused on models 3 and 5, characterized by $M_\bhb = 10^7\Ms$ and $e_\bhb=0.5$ and both $e_\gc=0.5$ and $e_\gc=0.9$. We investigated the effect of these two GCs on the galactic nucleus following a three step procedure:
\begin{enumerate}
\item we simply mixed up the snapshots of the two simulations, rescaling both of them in the BHB centre of mass to ensure a correct overlapping;
\item we randomized their orbits through a rotation along the three axis;
\item we duplicated $n$ times each cluster, in order to study how the debris distribution changes at increasing GC number.
\end{enumerate}
Figure \ref{N6rot} shows the surface density map and radial profile in the case of $N_\gc = 2$, $4$ or $6$ subsequent GCs infall assuming the simplifications above.

The choice $N_\gc = 2$ serves as a comparison with the results obtained with model 21, where we simulated 2 spiralling GCs. The similarity between Figure \ref{smap} and top panels in Figure \ref{N6rot} suggests that our procedure is reasonable, although the outcomes are not perfectly overlapped. 

Increasing the number of GCs leads to a shallower surface density distribution, which becomes already flat at $N_\gc = 6$. This will make the detection of the corresponding NSC very difficult, because NSCs are usually detected through an excess in the observed surface brightness. Note that the surface density map shows a complex morphology characterized by the GCs core remnant and their debris sparsed around the trajectory. These features are expected to dilute after the GCs merged together. 

Our results nicely compare with the results provided by \cite{maureira18} \citep[but see also][]{goicovic18}, who modelled the evolution of multiple gaseous clouds impacting a BHB. Although the gas physical processes can significantly differ from simple stellar dynamics, the overall structure found by \cite{maureira18} after the clouds disruption is quite similar to our results, although the ``gaseous'' case leads also to evident differences. For instance, in some of their models \cite{maureira18} found the formation of two accretion discs around each SMBH, which are clearly impossible to form in our simulations. Also, they found several gas streaming intersecting the binary orbit, while in our case these features are almost absent. This is probably due to our choice of setting the GC pericentre at twice the BHB initial apocentre, $r_p = 2r_{a,{\rm BHB}}$, while in \cite{maureira18} sample gaseous clouds dive deeper into the BHB orbit, all having $r_p < 2 a_\bhb$.

Interestingly, it seems that a larger number of merger events lead to a cored central density, making hard to identify any NSC in the galactic centre and even masking the presence of an BHB. On the other hand, it seems that the core size extends roughly up to the BHB semi-major axis, thus giving an alternative possibility to spot the BHB presence from the observed surface density profile.

\section{Discussion}

\subsection{The distribution of compact remnants around the BHB}

\subsubsection{Tidal disruption of red giant stars}
Our models represent a simplified simulation of a GC orbiting a BHB placed in the centre of a massive galaxies. The galaxy gravitational field is taken into account as an external static potential, and the cluster is represented by $65$k single mass particles. Moreover, neither stellar evolution nor a detailed treatment for stellar close encounters have been included here. 

Although these limitations, we can make use of our simulations to make predictions on the possible distribution of compact stellar objects around the BHB.

Stars passing sufficiently close to a single SMBH can be tidally disrupted or, possibly, completely swallowed.
A tidal disruption event (TDE) occurs whenever the stars, with radius $R_*$ and mass $m_*$, passes near the SMBH closer to a typical radius, called tidal radius or Roche radius:

\begin{equation}
r_{\rm R} = \eta R_*\left(\frac{M_{\rm SMBH}}{m_*}\right)^{1/3},
\label{roche}
\end{equation}
with $\eta\sim 1$, depending on the stellar equation of state \citep{hills75,Merri13}.
TDEs are often followed by the emission of an X-ray flare with a time-scale of a few years. Nowadays, the detection of these strong signals represents a unique possibility to infer clues on the central SMBH mass and structure \citep{vinko15,kochanek16,yang16,metzger16}.

Main sequence stars mass and radius are linked by a simple power-law 
\begin{equation}
\frac{R_*}{{\rm R}_\odot} = \alpha\left(\frac{m_*}{{\rm M}_\odot}\right)^\beta,
\end{equation}
with $\alpha$ and $\beta$ depending on the stellar mass \citep{demircan91,gorda98}. 
 Substituting into Eq. \ref{roche} we find
\begin{equation}
\frac{r_{\rm R}}{r_{\rm S}} = 5.06\left(\frac{M_{\rm SMBH}}{10^7 ~\Ms}\right)^{-2/3} \alpha\left(\frac{m_*}{\Ms}\right)^{-1/3+\beta},
\label{rstar}
\end{equation}
being $r_{\rm S}$ the Schwarzschild radius.

Assuming $M_{\rm SMBH}=10^8$ M$_\odot$, Eq. \ref{rstar} implies that stars with mass smaller than $0.88$ M$_\odot$ are swallowed by the SMBH before the gravitational pull rips them apart (direct plunge), while heavier stars can give rise to a TDE. The light curves and spectra of TDEs can be used to shed light on the galaxy nucleus dynamics \cite{hayasaki18}.

Assuming stellar masses in the range $m_* = 1 - 4 \Ms$, corresponding to stellar radii $R_* = 1-3 ~{\rm R}_{\odot}$ for main sequence stars (MS), we found at most 1-2 TDE candidates in the whole simulations sample.
However, as shown above, the GCs typical inspiral time is generally larger than $\sim 0.1$ Gyr, thus suggesting that the majority of stars in the range $m_* = 1 - 4 \Ms$ are already out of their MS phase when reaching the centre.
If these stars are delivered during their red giant phase (RG), their radii can extend up to $10^2~{\rm R}_\odot$, thus facilitating their disruption. 

During this evolutionary phase, RG stars have a relatively low density and an extended envelope, thus the intense tidal forces to which they are subjected are expected to strip away their envelope, while the remaining He-rich core continues orbiting around the central object. 
In the case of a single SMBH, tidal heating and GW emission drive the core inspiral until it is ultimately tidally disrupted or it falls into the SMBH \citep{bogdanovic14}. The recently observed flare classified as PS1-10jh in a galaxy at redshift $z=0.1696$ around a single SMBH with mass $M_\bh\sim 2\times 10^6\Ms$ represents a possible prototype for such a kind of stellar disruption \citep{geszari12}. Indeed, the light curve and spectra of PS1-10jh suggest that the disrupted star was the tidally stripped core of a RG with initial mass $> 1\Ms$ evolved off the MS over a time comparable to the age of the galaxy stellar population ($\sim 5$ Gyr).
Additionally, part of the removed envelope falls back toward the SMBH, possibly emitting long-lasting flares that are expected to contribute to at least $\sim 10\%$ of the whole TDE rate of massive galaxies \citep{macleod12}.

The presence of a BHB can even boost the TDE rate  \citep{ivanov05,che09,che11,Li17,frag18}, while the number of RGs inhabiting the galaxy nucleus can substantially be altered by spiralling GCs \citep{min16}.

This motivated us to look for a possible population of TDEs originating from evolved stars in our models. 
In order to infer the number of disrupted RGs, we assume $m_* = 1 - 4 \Ms$ and $R_* = 10^2~ {\rm R}_\odot$ and calculate the number of stars satisfying the condition $r_p<R_R$. In the following, we refer only to models with $M_\bhb = 10^7\Ms$. 

\begin{figure}
\includegraphics[width=8cm]{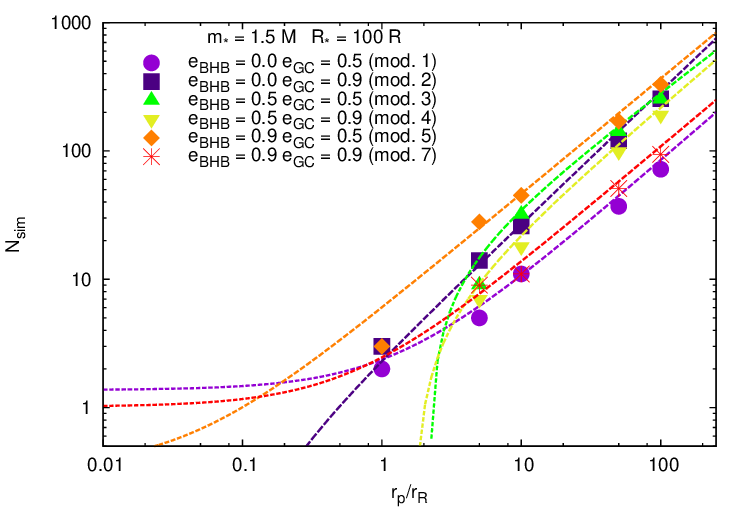}
\caption{Cumulative number distribution of stars pericentre for different models. Note that the pericentre is normalized to the Roche radius calculated for a star with mass $m_* = 1.5\Ms$ and radius $R_* = 100~{\rm R}_\odot$. }
\label{ntde}
\end{figure}

Figure \ref{ntde} shows the cumulative number of stars as a function of their pericentre, calculated after the complete GC disruption and normalized to the Roche radius. We found that the number of expected events is $N_{\rm sim} \simeq 1-10$ per GC. 

Using the results above, we can crudely estimate the rate of TDEs coming from disrupted clusters as
\begin{equation}
\Gamma_{\rm TDE} = N_{\rm sim} \times \frac{N_{\rm real}}{N} \times N_{\rm GC} \times f_{\rm DF} \times f_{\rm SE}\times T^{-1},
\label{ratetde}
\end{equation}
where $N_{\rm real}\sim 10^6$ is the actual number of stars populating a typical GC, $N_{\rm GC}$ is the total number of GCs, $f_{\rm DF}=0.16 $ is the fraction of GCs interacting with the BHB, $f_{\rm SE}= 0.03$ is the fraction of stars with masses $1 - 4 \Ms$, and $T=1-10$ Gyr is the typical time over which TDEs are expected to take place.
The total mass in GC in a galaxy is expected to be $\sim 1 \%$ of the total galaxy mass \citep[e.g.][]{gnedin14}, thus assuming a GC average mass $M_{\rm GC} = 10^6\Ms$ we can can calculate $N_{\rm GC}$ as
\begin{equation}
N_{\rm GC} = \frac{0.01M_g}{M_{\rm GC}}.
\end{equation} 
Assuming $M_g = 10^{11}\Ms$ and plugging numbers in Equation \ref{ratetde} we get a TDE rate for disrupted RG coming from infalling GCs 
\begin{equation}
\Gamma_{\rm TDE} = (0.7 - 7.2) \times 10^{-7} ~ {\rm yr}^{-1} \nonumber
\end{equation}

Equation \ref{ratetde} can also be used to calculate the TDE rate due to MS stars. To do this, we assume star masses between $0.5\Ms$ and $1\Ms$, which have a Roche radius larger than the BHB Schwarzschild radius for $M_\bhb = 10^7\Ms$ (see Equation \ref{rstar}). In this mass range, $f_{\rm SE} = 0.051$. The number of particles in our models having $r_p<r_R$ for MS stars, i.e. with $R_* \simeq 1{\rm ~R}_\odot$, is $N_{\rm sim} = 1$. This leads to a TDE rate $\Gamma_{\rm TDE} \simeq 1.2\times 10^{-7} {\rm yr}^{-1}$, comparable to the RG rate. Alike stars inhabiting the galactic nucleus for which the RG contribution to TDEs is expected to be $\sim 10\%$ \citep{macleod12}, our results suggest that the TDEs due to delivered stars is almost equally contributed by MS and evolved stars.

\subsubsection{Distribution of stellar-mass black holes}

Our models can be also used to constrain the distribution of stellar BHs delivered from the infalling GC in the BHB surroundings.
In order to do so, we firstly should estimate how many BHs are expected to form in the GC life-time.
Taking advantage from the BSE stellar evolution tool \citep{hurley00}, that we conveniently modified to include stellar wind prescriptions for massive stars provided by \cite{belckzinski10}, we found that the minimum mass for a stellar BH to form via single-star stellar evolution is $M_{\rm bht} \sim 18 \Ms$, in good agreement with recent mass estimates provided in literature (see for instance \cite{spera15}. According to a standard \cite{kroupa01} mass function, the number of stars with mass above  
$M_{\rm bht}$ is $N_{\rm bh} = 2.4\times 10^{-3}N_\gc$, being $N_\gc$ the total number of stars in the cluster.
For a mean stellar mass of $0.62\Ms$ and $M_\gc = 10^6$, this implies $N_{\rm bh} = 3871$ BHs. 
Promptly after their formation, BHs can receive a kick velocity as high as $\gtrsim 10^2$ km s$^{-1}$. 
However, currently is not completely clear the fraction of BHs receiving a kick sufficiently large to be ejected from the GC. Indeed, several studies pointed out recently that the ``BHs retention fraction'' is significantly larger than previously thought, reaching values significantly larger than $50\%$ \citep{morscher15,peuten16}.

In our simulations, we picked $N_{\rm bh}$ points in the GC initial conditions set and followed their evolution around the BHB.
We followed two different ``picking'' procedures:
\begin{itemize}
\item particles are picked randomly in the whole cluster;
\item particles are picked randomly within the cluster core.
\end{itemize}
We labelled each selected particle as a BH, and calculated the
global properties of such mock group.
This choice allow us exploring two extreme possibilities: i) BHs are still not segregated during the flyby, or ii) the BHs are strongly confined within the inner GC regions.
Possible differences between these two scenarios would provide interesting insights related to the stellar BHs distribution around heavy BHB which are potentially LISA\footnote{\url{https://www.lisamission.org/}} \citep{seoane07,eLISA13,amar17} or TianQin \citep{tianqin16} events.

\begin{figure}
\centering
\includegraphics[width=8cm]{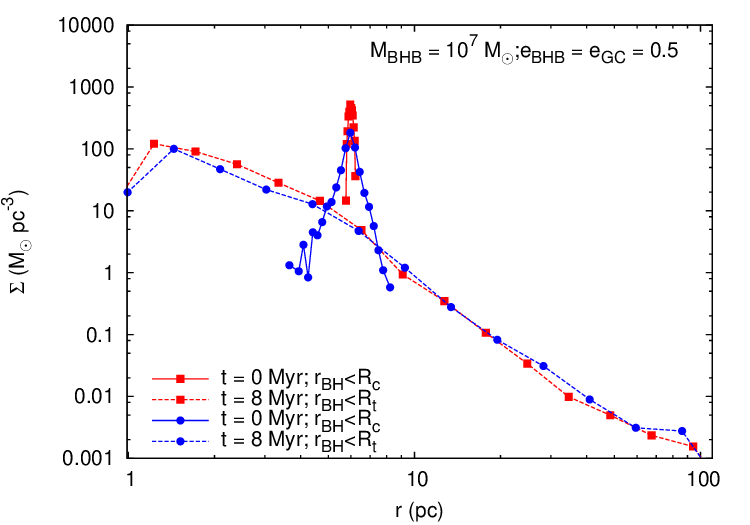}\\
\caption{BH density distribution for model $M_\bhb = 10^7\Ms$, $e_\bhb = e_\gc = 0.5$, assuming that the all the BHs were segregated within the cluster core radius (red squares) or not (blue points). Straight lines identify the initial $\rho_{\bh}$, while dashed lines show how it evolved after the GC disruption. The reference frame is assumed to be centred in the BHB centre of mass.}
\label{f13}
\end{figure}

As shown in Figure \ref{f13}, the BHB intense tidal field affects strongly the BH distribution, washing out any evidence of segregation after the GC disruption. Hence, it seems hard to infer any information on the BH initial distribution based on their distribution around a BHB.
Interestingly, this seems quite independent on the GC and BHB initial eccentricity.

\begin{figure}
\includegraphics[width=8cm]{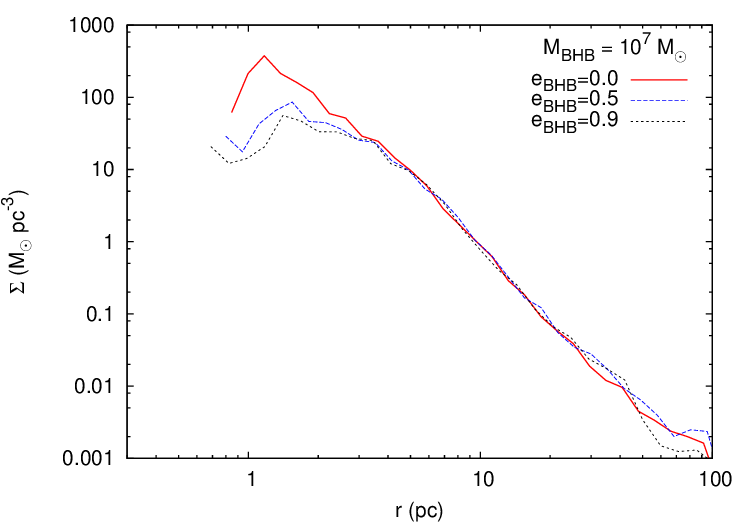}
\caption{Stellar BHs density profile, $\rho_\bh$, at varying $e_\bhb$, in the case of $M_\bhb=10^7\Ms$ and $e_\gc=0.5$.}
\label{f15}
\end{figure}

We found that a fraction $f_{\rm BH,ej} = 0.084 \pm 0.005$ is ejected from the clusters with average velocities $v_{\rm av} = 1568 \pm 1396$ km s$^{-1}$, with maximum velocities up to $v_{\rm ej} = 4200$ km s$^{-1}$. These calculations are based on BHs reaching a distance larger than 500 pc at time $T/T_\bhb = 5\times 10^3$, the velocities are calculated at the same time.
The BH distribution slightly changes depending on the BHB eccentricity, as shown in Figure \ref{f15}. Smaller $e_\bhb$ values leads to steeper $\rho_\bh$, while a BHB moving on a nearly radial orbit leads to a BHs distribution characterized by a core extending up to 5 times the initial BHB separation.

On one hand our results suggest that the distribution of compact remnants around the BHB can tell something about its orbital properties. On the other hand, the BHB tidal field seems dominating completely the GC final distribution, which loses memory of any previous segregation. Nearly $10\%$ of the BHs delivered by the GC are ejected at high velocities from the galactic bulge, while 
$50\%$ remain bound to the BHB. The remaining BHs can be either bound to the bulge, moving within the inner 500 pc, or high-velocity BHs that did not escape yet from the galactic nucleus.

From an observational perspectives, stellar BHs and other compact remnants are useful tools to constrain the origin of the nuclei they inhabit. 
Low mass X-ray binaries, millisecond pulsars and cataclysmic variables represent the ideal candidates to enlighten the BHB surroundings and provide useful insights on the properties of massive galactic nuclei.
They can be connected to the ``dry'' scenario for nuclear clusters formation and can be used to constrain the properties of their parent clusters \citep{brandt15,abbate17,ASK17,fragangn18,fpb18,fragione17b}. 

Hence, our results are twofold: on a side, they allows to infer the number of stellar BHs possibly gathered around the BHB, which can outnumber the population of galactic BHs, depending on the galaxy formation history; while on the other side they suggest that, differently from galaxies hosting a single SMBH, it is quite hard to infer the parent GC dynamical status (pre-infall) from the final debris distribution.

\subsection{BHB coalescence}

The BHB long-term evolution will be determined by a combination of two mechanisms: the continuous interactions with stars moving within the BHB loss-cone and 
GWs emission, which causes energy and angular momentum loss.

Stars passing within the loss-cone extract energy from the binary, forcing it to shrink, while tidal perturbations induced by the cluster debris might cause a variation in the binary eccentricity, which in turn can affect the amount of angular momentum carried away from GWs and its further hardening.

This complex picture is formalized through the  coupled system of differential equations \citep{quinlan96,sesana06}:

\begin{eqnarray}
\frac{{\rm d}a}{{\rm d}t} &=& \frac{{\rm d}a}{{\rm d}t}\Big|_{\rm SI}  + \frac{{\rm d}a}{{\rm d}t}\Big|_\gw,\\
\frac{{\rm d}e}{{\rm d}t} &=& \frac{{\rm d}e}{{\rm d}t}\Big|_{\rm SI} + \frac{{\rm d}e}{{\rm d}t}\Big|_\gw,
\label{eqGua}
\end{eqnarray}

whose solution allows to describe the SMBH secular evolution.

The first term in both equation represents the variation due to stellar interactions (SI), while the second term is due to GW emission.

The semi-major axis evolution due to stellar encounters can be expressed as 
\begin{equation}
\frac{{\rm d}a}{{\rm d}t}\Big|_{\rm SI} = -sa^2(t),
\label{EqSe1}
\end{equation}
provided that the binary hardening rate $s = {\rm d}a/{\rm d}t$ is nearly constant, while the GW term is given by \citep{peters63}
\begin{equation}
\frac{{\rm d}a}{{\rm d}t}\Big|_\gw = -\frac{64\beta}{5}\frac{F(e)}{a^3},
\label{EqSe2}
\end{equation}
with 
\begin{eqnarray}
F(e)  &=& (1 - e^2)^{-7/2}(1 +73/24 e^2 + 37/96e^4);\\
\beta &=& G^3/c^5 M_1M_2(M_1+M_2).
\end{eqnarray}

The eccentricity evolution regulated by stellar encounters and GW emission, instead, is given by 
\begin{eqnarray}
\frac{{\rm d}e}{{\rm d}t}\Big|_{\rm SI} &=& \frac{K}{a}\frac{{\rm d}a}{dt}\\
\frac{{\rm d}e}{{\rm d}t}\Big|_\gw      &=& -\frac{304\beta}{15}\frac{e G(e)}{a^4},
\label{eqE}
\end{eqnarray}
where 
\begin{equation}
G(e) = (1-e^2)^{5/2}\left(1+\frac{121}{304}e^2\right),
\end{equation}
 and $K$ is the eccentricity growth rate \citep{quinlan96}, which can be defined as 
\begin{equation}
K = \frac{{\rm d}e}{{\rm d}(\ln (1/a))}.
\end{equation}

Integrating Equations \ref{eqGua} requires a detailed knowledge of the hardening rate $s$ and the eccentricity growth rate $K$. 
Both these quantities depend on the properties of the environment in which the binary is embedded.
\cite{mikkola92} and later studies \citep{quinlan96,sesana06} parametrized the hardening rate through the adimensional $H(a_\bhb)$ hardening rate, being
\begin{equation}
s = H(a_\bhb) \frac{G\rho}{\sigma},
\end{equation}
with $\rho$ and $\sigma$ the density and velocity dispersion calculated within the BHs influence radius. Using scattering experiments, \cite{sesana06} provided a useful fitting formula for $H(a_\bhb)$
\begin{equation}
H(a_\bhb) = A(1+a/a_0)^k,
\end{equation} 
where the fitting parameters $A$, $a_0$ and $k$ depends on the BHB mass and mass ratio. In a spherically symmetric nucleus, the BHB interactions efficiently deplete the stars reservoir, emptying the so-called ``loss-cone'', driving the BHB into a regime where the energy loss due to the continuous scattering becomes inefficient. As a consequence, the binary stalls and $s$ tends quickly to zero. On the other hand, this process is avoided in merging galaxies, since the asymmetric star distribution resulting from the merger allows the loss-cone replenishment through centrophilic orbits, maintaining the binary in the full loss-cone regime and avoiding the stalling \citep{khan12,preto11,gualandris12}.

Our simulations can be considered, in principle, as a scaled version of a minor merger events, since the disrupting star cluster distributes its debris around the BHB, driving the formation of a strongly asymmetric star distribution. 
Star clusters debris help in sustaining the full loss-cone regime, a process that recently have been proposed to describe the possible evolution of a SMBH-IMBH pair inhabiting the Milky Way centre in its early life \citep{mastrobuono14,ASG16,frlgk18,fragk17}.

Moreover, orbitally segregated GCs are ideal sources to replenish the loss-cone over time. Indeed, in a realistic galaxy model, GCs will have different inspiral times and will interact with the BHB at different stages of the BHB evolution. 
Under the simplest assumption of a GCs-BHB interactions rate such to sustain the full loss-cone regime, we investigate the BHB long-term evolution to determine under which condition the BHB merge.

We find in some cases that the GCs-BHB interactions leads to a hardening rate nearly constant, similarly to what observed for non-spherical simulations of BHB in post-merged galaxies \citep{preto11,khan12,gualandris12,khan16}.

Even the $K$ parameter can be parametrized through fitting formulae developed through scattering experiments \citep{sesana06,berentzen09,preto11}
\begin{equation}
K = e(1 - e^2)^{c_0} (c_1 + c_2e),
\label{scat}
\end{equation}
where $c_i$ are parametrizing constants \citep{quinlan96}.
However, as we shown in the following, the $K$ time variation calculated directly from the $N$-body models is hard to fit with such smooth function \citep[see also][]{wang14}.
This is due to its dependence on the ratio between the $e$ and $1/a$ time derivatives, which assume relatively small values, as we show in the following, and their ratio can vary significantly even on a quite short time-scale. 

Moreover, it must be noted that Equation \ref{scat} is obtained under the assumption that the BHB sits in a homogeneous sea of intruders moving on unbound orbits, while in our models the system is highly anisotropic and the perturbing stars move on similar orbits, inherited from the GC orbital configuration.

Our calculations suggest that, after the cluster disruption, the BHB hardening rate evolves toward a nearly constant value, while the eccentricity derivative is characterised by a nearly zero average value.
Hence, in order to provide a qualitative picture of the BHB long-term evolution, we integrated Equations \ref{eqGua} tuning the $s$ value in such a way to reach agreement between the semi-analytic solution and the direct $N$-body simulation. 

\begin{figure}
\centering
\includegraphics[width=8cm]{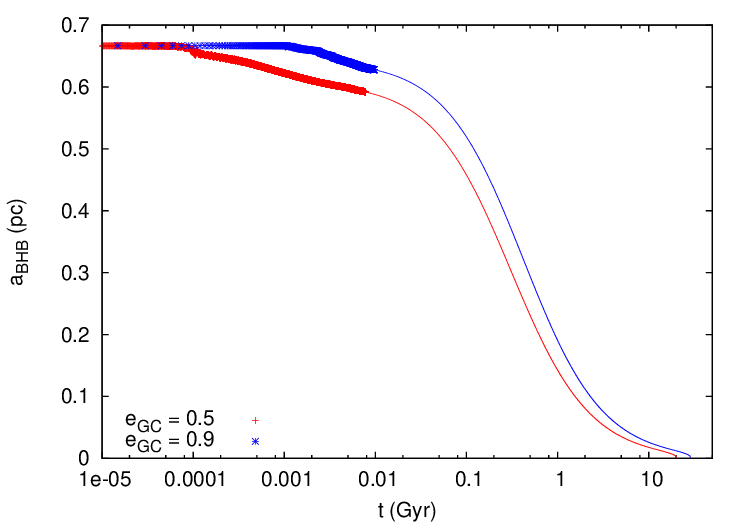}\\
\includegraphics[width=8cm]{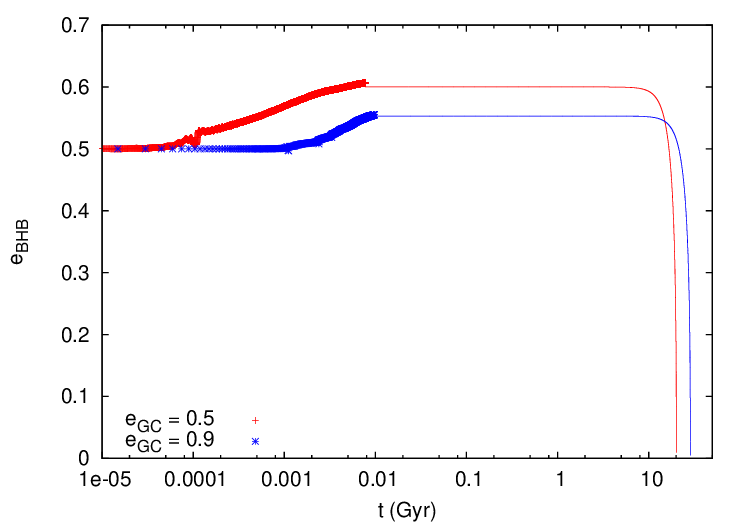}\\
\caption{Long term evolution of the BHB semi-major axis (upper panel) and eccentricity (lower panel) for two different models with $M_\bhb = 10^7\Ms$ and $e_\bhb=0.5$.}
\label{BHBm1}
\end{figure}

Figure \ref{BHBm1} shows a comparison between two different models with $e_\bhb = 0.5$ and different $e_\gc$. We note that when the GC moves on a nearly radial orbit its impact on the BHB evolution is less effective, determining a slower decrease of $a_\bhb$ and increase of $e_\bhb$. 

\begin{figure}
\centering
\includegraphics[width=8cm]{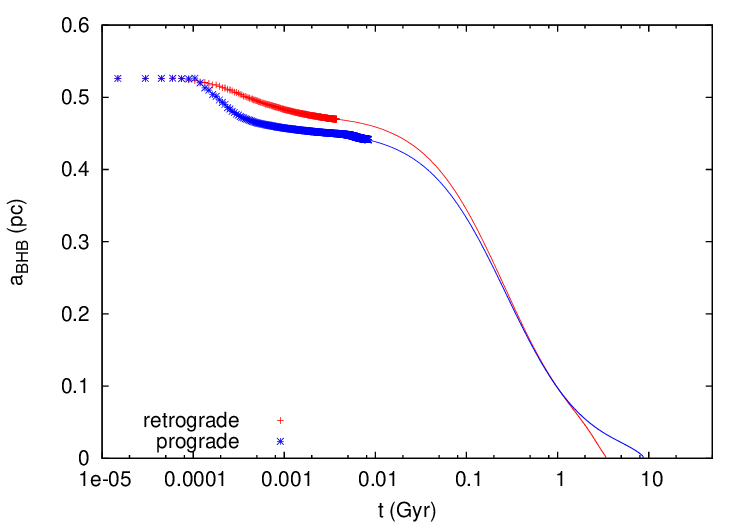}\\
\includegraphics[width=8cm]{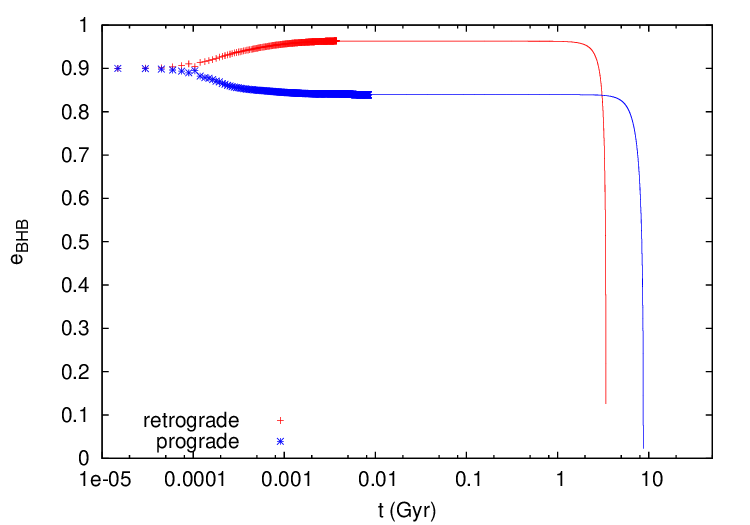}\\
\caption{As in Figure \ref{BHBm1}, but here we compare a prograde and retrograde cases characterized by $M_\bhb = 10^7\Ms$ and $e_\bhb=0.9$ and $e_\gc=0.5$.}
\label{BHBm2}
\end{figure}

The coalescence process changes significantly depending on the mutual inclination of the GC and the BHB orbits. 
Figure \ref{BHBm2} shows how the BHB parameters evolve for a prograde and retrograde cases. 
Initially, the BHB hardening is faster in the prograde case, as expected from the previous analysis. However, due to the different eccentricity in the two cases, on longer time-scales the GW emission becomes more efficient for the retrograde case, that eventually hardens faster and coalesce before the prograde one.
Note that the merging time-scale in the retrograde case is an order of magnitude smaller than in the prograde configuration, thus confirming the importance of the orbital inclination of the infalling GC. 

Studying the long-term evolution of BHB through detailed scattering experiments, \cite{sesana11} showed that if the number of stars moving on a prograde orbit exceeds the 70\%, the BHB tends to circularize due to the tidal effects arising from such interactions. 
This is evident in our simulations, as shown in the bottom panel Figure \ref{BHBm2}. The GC debris collective motion represents a large population of stars moving on a prograde (or retrograde) orbit, depending on the GC initial inclination, thus inducing an efficient eccentricity decrease (or increase).

Assuming $K=0$ is a rather strong assumption, since leads to neglect the eccentricity increase driven by retrograde stars.
To highlight the role played by the choice of $K$, we used either $K=0$ and $K$ calculated through Equation \ref{scat} for model No. 3 ($M_\bhb = 10^7\Ms,~ e_\bhb = e_\gc = 0.5$), using the coefficient values provided by \cite{quinlan96} for an equal mass binary. 

\begin{figure}
\centering
\includegraphics[width=8cm]{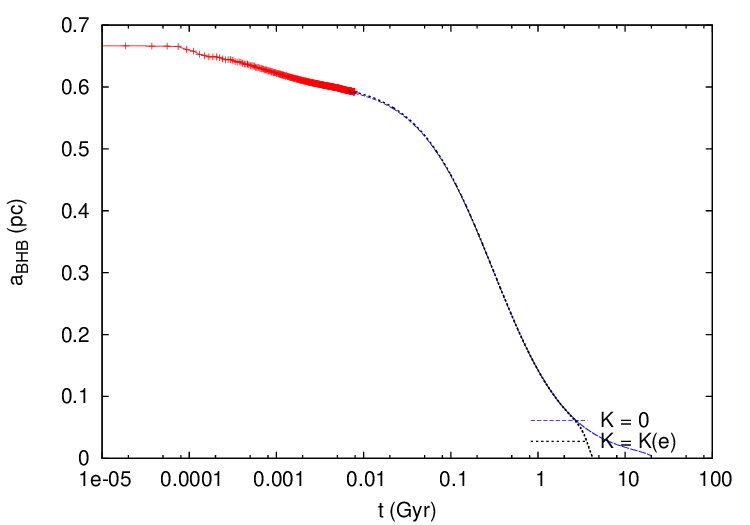}\\
\includegraphics[width=8cm]{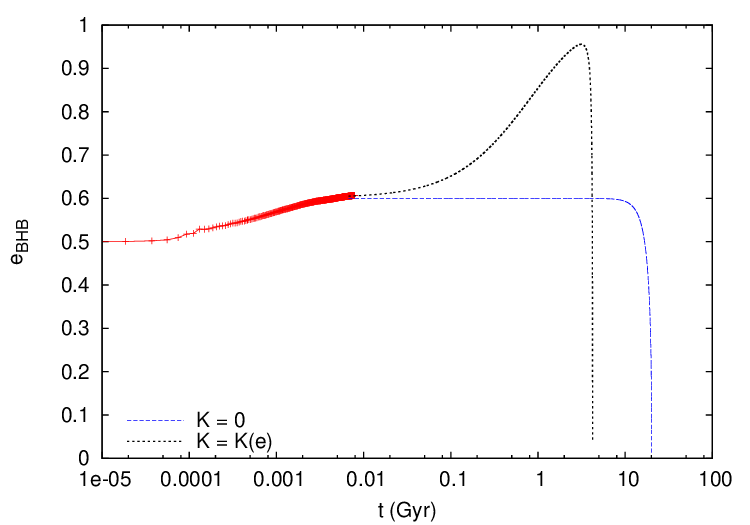}\\
\caption{Semi-major (top panel) and eccentricity (bottom panel) time evolution for model No. 3, assuming $K = 0$ (blue dashed line) or calculating $K$ through Equation \ref{scat} (black dotted line). The BHB mass considered here is $M_\bhb = 10^7 \Ms$, while the BHB and GC have same eccentricity, $e_\bhb = e_\gc = 0.5$.}
\label{Kanalysis}
\end{figure}

When including a proper treatment for the eccentricity evolution, the time needed for the BHB coalescence reduces by a factor $\sim 4-5$, as shown in Figure \ref{Kanalysis}. 

\begin{figure}
\centering
\includegraphics[width=8cm]{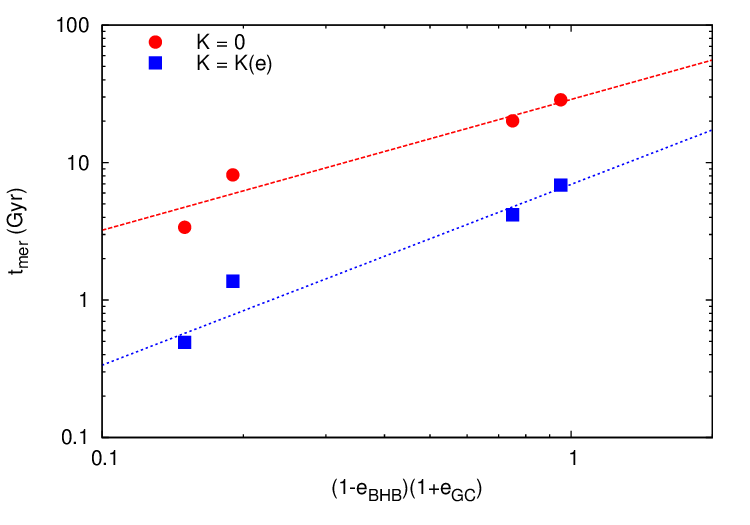}
\caption{Coalescence time as a function of the combined BHB and GC eccentricities. Different symbols correspond to different choice to calculate $K(e)$.}
\label{BHBm3}
\end{figure}

Comparing all the models in which the coalescence occurs within a time-scale $O(10 Gyr)$, we found that the merging time depends on the product $(1-e_\bhb)(1+e_\gc)$, as shown in Figure \ref{BHBm3}. 
We fitted the relation with a simple powerlaw $t_{\rm mer} = A[(1-e_\bhb)(1+e_\gc)]^\beta$.

The best fit parameters are $A= 28\pm 2$ Gyr and $\beta =  0.95\pm 0.19$ in the case $K=0$, and $A= 6.9\pm 0.7$ Gyr and $\beta =  1.32\pm 0.36$ if we assume $K = K(e)$. Table \ref{tabmerg} summarizes the merger time calculated in both the cases.

These results suggest that the merging time scales almost linearly with this combination of the GC and BHB eccentricities.
Clearly, a larger number of models would be required to confirm and refine the fitting formula in terms of BHB mass and initial semi-major axis, which here are not taken into account. 

Since the maximum value possible for $[(1-e_\bhb)(1+e_\gc)] = 2$, i.e. both the BHB and GC moves on a circular orbit, we can put an upper limit on the BHB merging time-scale $t_{\rm mer}\simeq 2A = 56$ Gyr ($K=0$), or $t_{\rm mer} \simeq 14$ Gyr ($K=K(e)$), in the case of an equal mass binary with total mass $M_\bhb=10^7\Ms$ and initial apocentre $r_a = 1$ pc.

\begin{table*}
\caption{}
\centering{Merging time-scales for several models assuming $K=0$}
\begin{center}
\begin{tabular}{ccccccccccc}
\hline
ID & i & $M_\bhb$ & $e_\bhb$ & $e_\gc$ & $t_0$ & $a_0$ & $e_0$ & $t_{\rm mer,0}$ & $t_{\rm mer,0}$ & $t_{\rm opt,0}$\\ 
   & ($\circ$) &$10^7M_\odot$ & & & NB & pc & & $10^7$ NB & Gyr & Gyr \\
\hline
3 &$180$ & $1.0$ & $0.5$ & $0.5$ & $5169$  & $0.592$ & $0.607$ & $ 1.35$  & $20$  & $4.15$\\
4 &$180$ & $1.0$ & $0.5$ & $0.9$ & $5714$  & $0.629$ & $0.553$ & $ 1.92$  & $28$  & $6.71$\\
5 &$180$ & $1.0$ & $0.9$ & $0.5$ & $2003$  & $0.471$ & $0.963$ & $ 0.227$ & $3.4$ & $0.49$\\
6 &$0$   & $1.0$ & $0.9$ & $0.5$ & $4996$  & $0.442$ & $0.600$ & $ 1.95$  & $29$  & $38.2$\\
7&$180$  & $1.0$ & $0.9$ & $0.9$ & $5997$  & $0.496$ & $0.945$ & $ 0.546$ & $8.1$ & $1.36$\\
\hline
\end{tabular}
\end{center}
\label{tabmerg}
\end{table*}

Promptly after the merger, the resulting SMBH is expected to get a recoil kick due to anisotropy GW emission \citep{herrmann07a,herrmmann07b}, whose amplitude depends on the total spin and the ratio between the BHB reduced and total mass $\eta_\bhb = \mu_\bhb/M_\bhb$ For an equal mass BHB for maximally rotating SMBHs with aligned spins, the recoiling velocity can be as high as $2000$ km s$^{-1}$ \citep{schnittman07,lou08,lou10,lou11}.
In such a case, the resulting SMBH will be ejected from the galactic centre while, at lower kicks, the SMBH displacement can carve the galaxy nucleus, leading to the formation of an extended core \citep{campanelli07}.

\subsection{The fate of the Milky Way and Andromeda SMBHs}
\label{milkomedasec}

Over the next $\sim 4$ Gyr, the Milky Way is expected to collide with its twin Andromeda, leading to the formation of Milkomeda \citep{Cox07}. In consequence of this catastrophic event, the two SMBH hosted in both galaxies are expected to bind together and form an unequal mass BHB. While the MW SMBH (Sgr A*) has a mass $M_{\rm Sgr A*} = 4.5\times 10^6\Ms$ \citep{ghez08,gillessen09,schodel14}, the Andromeda SMBH is much heavier, $M_{\rm M31} =  10^8\Ms$ \citep{bender05}. 
The galaxy merger is expected to trigger a burst of star formation, possibly leading to massive star clusters that eventually undergo orbital segregation and impact over the newly formed BHB.
The solar system and the Earth will be pulled on a wider orbit, extending outside 20-40 kpc from the Milkomeda centre and likely will inhabit one of the long tidal tails formed during the merger. Milkomeda is expected to be a massive elliptical galaxy with properties similar to our galaxy model, thus representing a perfect example to be explored with our simulations.

The Milkomeda BHB represents a potentially interesting BH pair, being characterized by a mass ratio of only $M_{\rm Sgr A*}/M_{\rm M31} = 0.05$. 

The merging phases of the MilkyWay-Andromeda system will be very complex. Promptly after the galaxy collision, the two SMBHs will spiral due to dynamical friction, eventually binding in a tight pair that will shrink due to the continuous interaction with field stars and gas.

The arrival of an infalling GC can boost the BHB evolution, as shown in the previous sections, driving it rapidly toward coalescence.

In order to determine whether future astronomers will have the possibility to witness from a preferential location a BHB merger, we simulated the evolution of the Milkomeda BHB, assuming $e_\bhb=e_\gc=0.5$ and setting the infalling GC pericentre $r_p = 2$ pc. The results can be directly compared with our model 12 in Table 1, characterized by an equal mass binary with $M_\bhb = 10^8\Ms$.

\begin{figure}
\centering
\includegraphics[width=8cm]{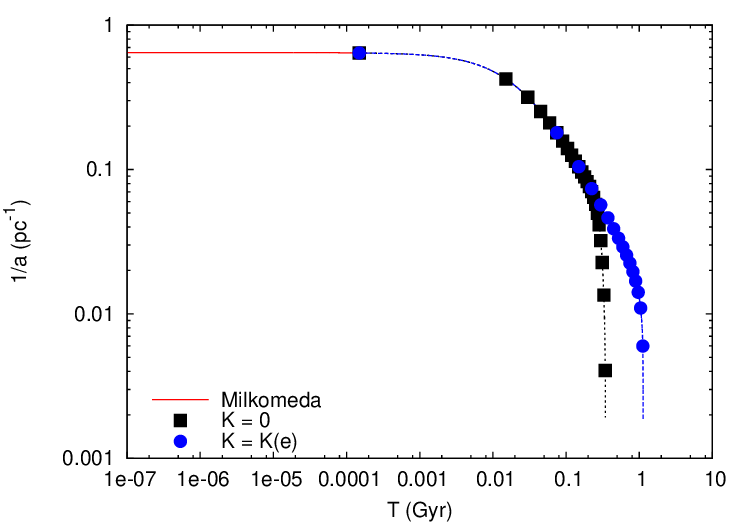}\\
\includegraphics[width=8cm]{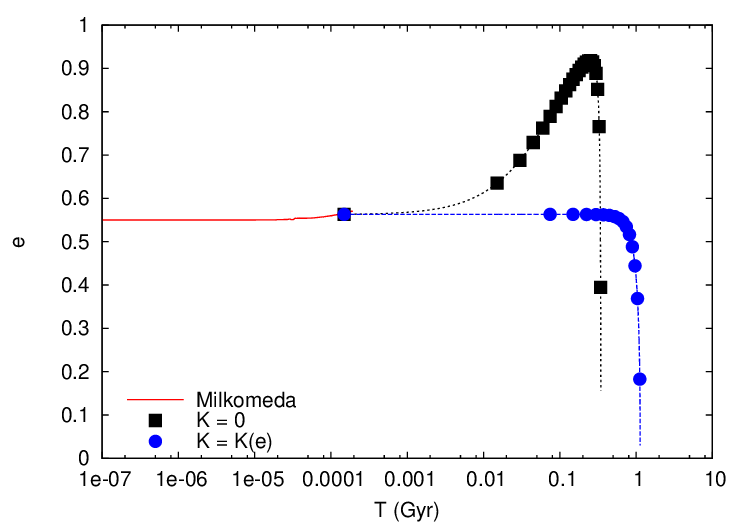}\\
\caption{Inverse semi major axis and eccentricity of the simulated Milkomeda BHB. The red straight line labels the simulation data, the extrapolation performed solving Eq. \ref{eqGua} is shown assuming $K = 0$ (blue dashed line), or $K=K(e)$ (black dotted line).}
\label{final}
\end{figure}

We carried out our model up to 30 Myr, following the subsequent evolution of the BHB semi-major axis and eccentricity following the treatment described in Sect. \ref{BHBMER}. Our results are shown in Figure \ref{final}.

Due to the small ratio between the BHB reduced and total mass, $\eta_\bhb = 0.041$, the post-merged SMBH is expected to receive a small kick $v_k \sim 10-50$ km s$^{-1}$ even for maximally rotating SMBHs \cite{schnittman07}, thus suggesting that the Milkomeda will continue its life as a massive elliptical galaxy hosting a central 
SMBH.

\section{Conclusions}
This paper presents accurate simulations of the interaction of a massive globular cluster with a super massive binary black hole orbiting in the host galaxy center.
The main results of our work can be summarized as follows:

\begin{itemize}
\item a close fly-by between a GC and a BHB can alter notably the BHB semi-major axis and eccentricity, depending on the orbital configuration. The rate at which the BHB semi-major axis and eccentricity vary over time depends mostly on the BHB initial orbital parameters. The BHB hardening rate increases at decreasing values of the BHB eccentricity, most likely due to the fact that in our models a lower eccentricity correspond to a smaller BHB binding energy. In general, we find that in retrograde configurations the larger the initial eccentricity the faster the eccentricity growth, thus the shorter the coalescence time-scale.

\item A substantial increase of the eccentricity is a common feature of almost all the retrograde models. As opposed to this, close GC-BHB interactions occurring in a prograde configuration lead the BHB to circularize. Also the BHB hardening efficiency depends on the configuration. Our models suggest that the semi-major axis reduction in a prograde configuration can be $\sim 50\%$ larger than for retrograde models. 

\item the GC is almost destroyed by the close encounter with the BHB, leaving debris around the BHB with a distribution in the angular momentum space quite different for retrograde and prograde orbits. This can have implications for detecting signature of recent GC-BHB interactions in galactic nuclei;

\item during the flyby, a sizeable number of stars are ejected away from the galactic nucleus, reaching velocities up to $\sim 1000-3000$ km s$^{-1}$. These stars can be detected as high-velocity stars (HVSs) in the galactic halo, thus suggesting that the presence of a large number of HVSs might be connected with the presence of a massive BHB in the galactic centre;

\item a small fraction of GC stars pass near enough to the BHB to be tidally disrupted, giving a TDE rate $0.7 - 7.2\times 10^{-8}$ $yr^{-1}$, an order of magnitude smaller than the estimates provided for galaxies hosting single SMBHs. These events are expected to occur in bursts, whose time interval depends on the GC infall rate;

\item the distribution of compact remnants, either BHs, NSs or WDs, delivered by the infalling cluster around the BHB depends on the BHB eccentricity. For nearly a circular BHB, the distribution is characterized by a steep rise, while for the eccentric models the distribution is nearly flat, with a ``core'' extending up to $\sim 3$ pc if $M_\bhb = 10^7\Ms$;

\item we demonstrated that, in some of the cases studied, when $M_\bhb = 10^7\Ms$,
the long term evolution of the GC debris around the BHB can drive the BHB coalescence within a Hubble time. In particular, we have found a simple relation connecting the GC and BHB eccentricities to the merger time-scale;

\item the GCs debris may leave an observational fingerprint in the galaxy surface density profile. The properties of such fingerprints depend primarily upon the GCs initial conditions. Our models suggest that several infalling clusters can give rise to a peculiar trend in the surface density and lead to nuclear cluster or nuclear disc-like features;

\item we explored the possibility that the Milky Way and Andromeda BHs will bind and coalesce in the nucleus of the resulting galaxy ``Milkomeda". We find that a massive perturber moving on a mildly eccentric orbit can boost the BHB eccentricity increase, reducing significantly the BHB lifetime. If such kind of perturbation is supplied by continuous  interactions with several perturbers, thus sustaining the BHB hardening rate and the eccentricity growth rate, our model suggests that the BHB merger will occur in $\sim 5$ Gyr from today. 
\end{itemize}

Although exploring a limited portion of the phase-space, our results suggest that spiralling GCs can significantly impact the BHB evolution. Using these simulations as a guideline, in future works we will develop a new set of models that will cover regions of the phase-space still unexplored. As a natural continuation of this work, we will study models at varying the GC structural parameters, an aspect that we discussed only partly in this work. Many GCs can further accelerate the BHB evolution, as demonstrated in one of our supplementary models, thus making the study of multiple GCs orbiting the SMBH an appealing system worth of further investigation. Simulating both the infalling GC and galaxy stars, at least those in the galaxy nuclear regions, will represent a crucial step forward for this kind of models since it will allow to properly follow the orbital evolution of the GC and the combined effect of GC and galaxy stars on the BHB evolution. 

\section*{Acknowledgement}
The authors acknowledge the anonymous referee, whose meaningful comments and suggestions allowed us to greatly improve the earlier version of the manuscript.
The authors thank Andreas Just for useful discussions that helped to improve the quality of the manuscript, and acknowledge Alessia Gualandris for having provided her numerical code to integrate the BHB long-term evolution.
The authors gratefully acknowledge the Gauss Centre for Supercomputing e.V. (www.gauss-centre.eu) for funding this project by providing computing time through the John von Neumann Institute for Computing (NIC) on the GCS Supercomputers JURECA and JUWELS at J\"ulich Supercomputing Centre (JSC). 
MAS, PB, MS, RS and RCD acknowledge the Sonderforschungsbereich SFB 881 "The Milky Way System" (subproject Z2) of the German Research Foundation (DFG) for the financial support provided. MAS also acknowledges the Alexander von Humboldt foundation and the Federal Ministry for Education and Research, which provided financial support to the research project ``The evolution of black holes at all the scales''.
This work was done in the footsteps of the  ``MEGaN project: modelling the evolution of galactic nuclei'', funded by the University of Rome Sapienza through the grant 52/2015.
PB, MS, and RS acknowledge support by Volkswagen Trilateral Project 90411, Dynamical Mechanisms of Accretion in Galactic Nuclei.
GF is supported by the Foreign Postdoctoral Fellowship Program of the Israel Academy of Sciences and Humanities. GF also acknowledges support from an Arskin postdoctoral fellowship and Lady Davis Fellowship Trust at the Hebrew University of Jerusalem.
We acknowledge support from the Strategic Priority Research Program (Pilot B) ``Multi-wavelength gravitational wave universe'' of the Chinese Academy of Sciences (No. XDB23040100).

\clearpage
\footnotesize{
\bibliographystyle{mnras}
\bibliography{ASetal2015}
}

\bsp	
\label{lastpage}
\end{document}